\documentclass[journal]{IEEEtran}
\usepackage{graphicx}
\usepackage{subfig}
\usepackage{pgfplots}
\usepackage[font=small]{caption}
\usepackage{subfiles}
\usepackage{float}
\usepackage{url}
\usepackage{multirow}
\usepackage{colortbl}
\usepackage{amsmath}
\usepackage{xcolor}
\usepackage{balance}
\usepackage{tikz}
\usetikzlibrary{tikzmark}
\usepackage{hyperref}
\pgfplotsset{width=10cm,compat=1.9}
\usepackage{url}

\date{May 2020}

\begin{document}
%\mainmatter

\title{Optimizing Temporal Convolutional Network inference on FPGA-based accelerators}

% \markboth{IEEE Journal on Emerging and Selected Topics in Circuits and Systems}{Carreras \MakeLowercase{\textit{et al.}}: Optimizing Temporal Convolutional Network inference on FPGA-based accelerators}

% \author{Marco Carreras\inst{1}, Gianfranco Deriu\inst{1}, Luigi Raffo\inst{1}, Luca Benini\inst{2}, Paolo Meloni\inst{1}}
% %\authorrunning{Carreras et al.}

% \institute{Universit\`{a} degli Studi di Cagliari, DIEE\\
% \email{marco.carreras@unica.it}\\
% \email{gianfranco.deriu@unica.it}\\
% \email{paolo.meloni@unica.it}}

%1\textsuperscript{st}

%\author{\IEEEauthorblockN{Marco Carreras\IEEEauthorrefmark{1}, Gianfranco Deriu\IEEEauthorrefmark{1}, Luigi Raffo\IEEEauthorrefmark{1}, Luca Benini\IEEEauthorrefmark{2} and Paolo Meloni\IEEEauthorrefmark{1}
%\IEEEauthorblockA{\IEEEauthorrefmark{1}\textit{DIEE},
%\textit{Universit\`{a} degli Studi di Cagliari}\\
%Cagliari, Italy \\
%\{marco.carreras, gianfranco.deriu, paolo.meloni, raffo\}@unica.it}}
%\IEEEauthorblockA{\IEEEauthorrefmark{2}\textit{Universit\`{a} degli Studi di Bologna}, Bologna, Italy and \\ \textit{ETHZ},
%Zurich, Switzerland \\
%luca.benini@unibo.it, lbenini@ethz.ch}
%}

\author{Marco~Carreras,
        Gianfranco~Deriu, Luigi~Raffo, Luca~Benini
        and~Paolo~Meloni% <-this % stops a space
\thanks{M. Carreras, G. Deriu, L. Raffo and P. Meloni are with the Department
of Electrical and Computer Engineering, Universit\`{a} degli Studi di Cagliari, Cagliari,
Sardinia, 09123 Italy (e-mail: marco.carreras@unica.it, gianfranco.deriu@unica.it, paolo.meloni@unica.it, raffo@unica.it).}% <-this % stops a space
\thanks{L. Benini is with Universit\`{a} di Bologna, , Bologna, Italy and ETHZ, Zurich, Switzerland (e-mail: luca.benini@unibo.it and lbenini@ethz.ch).}% <-this % stops a space
\thanks{This work has received funding from the European Union’s Horizon 2020
Research and Innovation Programme under grant agreement No. 780788.}}

\maketitle

\begin{abstract}
Convolutional Neural Networks are extensively used in a wide range of applications, commonly including computer vision tasks like image and video classification, recognition and segmentation. Recent research results demonstrate that multi-layer (deep) network involving mono-dimensional convolutions and dilation can be effectively used in time series and sequences classification and segmentation, as well as in tasks involving sequence modelling. These structures, commonly referred to as Temporal Convolutional Networks (TCNs), have been demonstrated to consistently outperform Recurrent Neural Networks in terms of accuracy and training time \cite{bai2018empirical}.
While FPGA-based inference accelerators for classic CNNs are widespread, literature is lacking in a quantitative evaluation of their usability on inference for TCN models.
In this paper we present such an evaluation, considering a CNN accelerator with specific features supporting TCN kernels as a reference and a set of state-of-the-art TCNs as benchmark. Experimental results show that, during TCN execution, operational intensity can be critical for the overall performance. We propose a convolution scheduling based on batch processing that can boost efficiency up to 96\% of theoretical peak performance. Overall we can achieve up to 111,8 GOPS/s and a power efficiency of 33,9 GOPS/s/W on an Ultrascale+ ZU3EG (up to 10x speedup and 3x power efficiency improvement with respect to pure software implementation).

\medskip
\end{abstract}

\section{Introduction}
The widespread use of Convolutional Neural Networks (CNNs), as a go-to solution for a wide range of AI-related problems, combined with the high computational load typically associated with their execution, has led researchers to extensively work on the development of hardware accelerators for CNN inference \cite{inmem}\cite{eyeriss}\cite{hyperdrive}\cite{CNN-P}. Availability of this kind of hardware is key in embedded use-cases involving near-sensor processing of data, according to the edge computing paradigm \cite{vestias2019survey}. Among the different solutions available in literature, an important role is played by FPGA-based architectures, and, in more detail, by solutions exploiting the cooperation between general-purpose processors and FPGAs available in modern All-programmable SoCs, e.g. Zynq and Zynq Ultrascale+ devices by Xilinx, that take profit from the efficient implementation of Multiply-And-Accumulate (MAC) operations on the large amount of DSP Slices available \cite{survey}.
CNNs have been employed for computer vision applications, such as image recognition \cite{img_rec_He_Zhang,imagenet_Krizhevsky,deepImage}, object detection \cite{deepFace} or video frame classification \cite{videoClass}. However, convolutional networks have also been extended to deal with time sequences. These CNN variations are usually indicated as Temporal Convolutional Networks (TCNs) \cite{lea2017temporal}. Multiple TCN architectures have been proposed, reaching impressive performance on tasks such as sentence classification \cite{sentClass}, speech recognition \cite{deepSpeech}, text understanding \cite{textUnde}, Natural Language Processing tasks \cite{NLP} and, more recently, machine translation \cite{machineTras}, audio synthesis \cite{wavenet}, language modeling \cite{lang_model} or signal sequence analysis in the healthcare domain, such as action detection \cite{3D_skeleton} or ECG classification \cite{ecg}.
Research work of Bai et al. \cite{bai2018empirical} demonstrates that the exploitation of a Temporal Convolutional Network for typical sequence modelling tasks often outperforms older and better-known Recurrent Neural Networks \cite{lstm} \cite{gru}. \newline
Thus, the rapidly increasing interest in TCNs pushes for investigating on acceleration for these networks. Especially in the embedded domain, where (near) real-time analysis of sequences of data samples acquired by sensors is a common case, accelerating this kind of workload on reconfigurable devices is a very appealing approach.  
In this work we explore the capabilities of a state-of-the-art CNN inference accelerator \cite{neu2017} specifically enriched to provide the flexibility needed in TCNs, to support freely selectable \emph{kernel sizes} and \emph{dilated} convolutions, with freely selectable \emph{dilation rates}. We focus on an implementation on low-cost and low-power all-programmable SoCs, more suitable for the integration of edge-computing and IoT processing nodes, considering two widely accessible devices in the Zynq and the Zynq Ultrascale+ families. 
Performance is evaluated over various benchmark TCNs, reporting on absolute execution time as well as on the efficiency of the execution with respect to the peak performance imposed by the device resources. 
We propose an optimization method relying on \emph{batch processing} to improve efficiency in cases where operational intensity is critical. We also assess the efficiency of the FPGA-based acceleration, comparing with software execution and with state of the art accelerators on bi-dimensional CNNs.

The reminder of this paper is organized as follows: Section \ref{sec:related} discusses the related work. Section \ref{sec:tcn_gen} analyses the TCN model. Section \ref{sec:ref_arch} gives an overview of the accelerator architecture taken as reference for this work. Section \ref{sec:CE} describes implementation details and strategies. The experimental results are presented in Section \ref{sec:hw_impl_eval} and Section \ref{sec:exp_res}. Conclusions are exposed in Section \ref{sec:concl}.
\section{Related Work}
\label{sec:related}
The landscape of FPGA-based accelerators for CNN is crowded and multifaceted \cite{guo2019dl}. Several approaches have been proposed in recent years, focusing both on the embedded domain and on architectures aimed to speed-up execution on cloud servers.
However, work on FPGA acceleration has mainly focused on classic CNN networks. %Literature proposes a significant amount of approaches that are more related to this paper, focusing on embedded FPGAs and All-programmable SoCs.

Yu et al. \cite{yu2019data} developed an FPGA acceleration platform that leverages a unified framework architecture for general-purpose CNN inference acceleration at a data center achieving a throughput comparable with the state-of-the-art GPU in this field, with less latency. This work exploits on-chip DSPs, on a Kintex KU115, arranged as supertile units (SUs), to overcome the computational bound and, together with dispaching-assembling model and broadcast caches, to deal with the memory bound.

Zhang et. al. \cite{Caffeine} proposed Caffeine, a hardware/software library to efficiently accelerate CNNs on FPGAs, leveraging a uniformed convolutional matrix multiplication representation targeting both computation-intensive convolutional layers and communication-intensive fully connected layers of CNN which maximizes the underlying FPGA computing and bandwidth resources utilization. Ma et. al. \cite{RTLCompiler} presented an RTL-level CNN compiler that generates automatically customized FPGA hardware for the inference tasks of CNNs from software to FPGA. The approach proposed by \cite{RTLCompiler} relies on a template accelerator architecture described in Verilog including all the main functions employed by CNNs such as convolutions, pooling, etc, which are automatically customized at design time to match the requirements of the target CNN model.
The proposed methodology is demonstrated with end-to-end FPGA implementations of complex CNN models such as NiN, VGG-16, ResNet-50, and ResNet-152 on two standalone Intel FPGAs, Stratix V and Arria 10, providing average performance up to 720 GOps.

These two frameworks provide huge performance gains when compared to state-of-the-art accelerators and general-purpose CPUs and GPUs. However, both leverage large FPGA devices such as Virtex7 and Arria 10, they mainly target server applications exploiting batching to improve memory access performance and bandwidth utilization.

This approach is less suitable for embedded applications where cheap and compact SoCs integrating embedded processors and FPGAs are desirable, and images have to be processed in real-time. %In this embedded domain, most recent works exploit the capabilities of Xilinx Zynq Z-7045 SoC, integrating a dual-core Cortex A9 processor operating up to 800 MHz and reconfigurable logic featuring 900 DSP slices.
In this domain, Venieris et. al. \cite{LatencyDriven} presented a latency-driven design methodology for mapping CNNs on FPGAs. As opposed to previously presented approaches mainly intended for bandwidth-driven applications, this work targets real-time applications where the batch size is constrained to one, relying on Xilinx high-level synthesis tools for mapping (i.e. Vivado HLS), demonstrated on relatively simple CNN such as AlexNet, and on a very regular one such as VGG16 featuring only 3$\times$3 kernels, providing a peak performance of 123 GOps. Other work focuses on a template-based approach based on programmable or customizable RTL accelerators proposed in architectures \cite{RTLCompiler}\cite{Snowflake}\cite{GoingDeeper}, more similar to the one that is used in this paper.

SnowFlake \cite{Snowflake} exploits a hierarchical design composed of multiple compute clusters. Each cluster is composed of four vectorial compute units including a vectorial MAC, vectorial max, a maps buffer, weights buffers and trace decoders. SnowFlake provides a computational efficiency of 91\%, and an operating frequency of 250 MHz (best-in-class for CNN accelerators on Xilinx Zynq Z-7045 SoC). However, although the vector processor-like nature of the accelerator is very flexible, delivering significant performance also for 1$\times$1 kernels, it prevents to fully exploit of spatial computation typical of application-specific accelerators, which leads to overheads due to load/store operations necessary to fetch weights and maps from the buffers. This is highlighted by the low utilization of the DSP slices available on the FPGA (i.e. only 256 over 900), and by the performance when executing end-to-end convolutional neural networks, which is lower than that of other architectures including the proposed one even though the operating frequency of the CNN engine is significantly higher.

Several approaches tackling FPGA architectures for image-processing CNN, have explored the reduction of the precision of arithmetic operands to improve energy efficiency. Although most of the architectures available in literature feature a precision of 16-bit (fixed-point)\cite{LatencyDriven, Snowflake, RTLCompiler} numerous reduced-precision implementations have been proposed recently, relying on 8-bit, 4-bit accuracy for both maps and weights, exploiting the resiliency of CNNs to quantization and approximation \cite{GoingDeeper}. 

Qiu et. al. \cite{GoingDeeper} proposed a CNN accelerator implemented on a Xilinx Zynq platform exploiting specific hardware to support 8/4 bit dynamic precision quantization, at the cost of 0.4\% loss of classification accuracy. 
Other extreme approaches to quantization exploit ternary \cite{Ternary} or binary \cite{FINN} neural-networks accelerators for FPGA. This approach significantly improves the computational efficiency of FPGA Accelerators, allowing to achieve performance level as big as 8 TOPS \cite{Ternary}.

Recent work by Rasoulinezhad et al. \cite{pir-dsp}, starting from the Xilinx DSP slices, proposed an optimized DSP block called PIR-DSP to efficient map 9, 4 and 2 bits data precision MAC operations. It is implemented as a parameterized module generator targeting both FPGAs and ASICs reaching an estimate run time energy decrease up to 31\% for a MobileNet-v2 implementation compared with a standard DSP mode.
Other works, like Wang et al. \cite{wang2020lutnet}, leverage FPGA LUT blocks as inference operators for Binary Neural Network (BNN) achieving up to twice area efficiency compared to state-of-the-art binarized NN implementation and against several standard networks models.
 
%These improvements are due to the 32-bit multipliers that can be replaced by simpler multiplexer and 2's complement operators, while bandwidth for loading weights can be reduced drastically, by 8 to 16 times if we compare with widely used 16-bit fixed point accelerators. The main issue related to binary and ternary accelerator is related to the training. 
While small networks like MNIST, CIFAR10, SVHN, GTSRB can reach good classification accuracy, the training is still a big challenge for larger networks such as VGG or ResNet \cite{Courbariaux2015a}. The usability of extreme quantization is also not demonstrated for TCN-related tasks, sequence classification and modelling. 

Probably the most powerful currently available FPGA-based acceleration engine is the proprietary one offered by Xilinx, which provides an integrated framework, called VitisAI \cite{VitisAI}, that helps designers in mapping CNNs on a templated soft IP called Deep Learning Processing Unit (DPU) \cite{DPU}. The DPU provides impressive performance on CNNs, using quantization and high clock frequency in DSP slices. Quantization is required and can be applied automatically using a dedicated tool included in VitisAI. DSP slices are clocked at very high frequency, using a \emph{DSP Double Data Rate (DDR) technique} \cite{DPU_guide}, which uses a 2x frequency domain to increase peak performance. However, the support for TCN is missing, since the DPU does not support arbitrary dilation, stride and kernel sizes and the the VitisAI quantization process does not support 1D convolutions.
To the best of our knowledge, there are no published FPGA-based accelerators tuned to speed-up inference for generic Temporal Convolutional Networks. 

As main novel contributions of this work, we propose:
\begin{itemize}
    \item an enriched architectural template supporting efficient TCN inference on FPGA;
    \item a methodology for the optimal execution/scheduling of data-transfers exploiting the specific sequence-based structure of data in TCNs
    \item a methodology for improving efficiency based on \emph{sample batching} (sequence buffering) 
    \item the first (to the best of our knowledge) experimental evaluation of the usability of FPGA-based acceleration for TCNs, based on different end-to-end benchmarks and two APSoC devices
\end{itemize}

\section{TCN model generalities}
\label{sec:tcn_gen}

A dilated convolution operation $F$ on element $s$ of a sequence \cite{bai2018empirical} can be defined as:

\begin{equation}
     F(s) = (x\ast_d f)(s) = \sum_{i=0}^{k-1} \ f(i)\cdot x_{s-d\cdot i}
     \label{eq:tcn_eq}
\end{equation}

where $x\in \mathbf{R}^n $ is a 1-D input sequence, $f:\{0,...,k-1\} \in \mathbf{R}$ is a kernel of size $k$ and $d$ is the $dilation\ rate$.

The sequence of samples that constitute the input of a TCN can be processed both off-line or in a real-time streaming. The second case is very useful in application cases requiring continuous analysis of the input sequence, e.g. aimed at the identification of specific events and/or at promptly closing the loop on data-triggered actuations. This implies that the sequence must be analyzed at  every time step, after being updated with a new sample. It is possible to identify the minimal sequence
%The latter implies that the sequence must be analyzed within every time step and updated with a new sample during the next step. The minimal sequence 
size to produce a valuable output sample as the \emph{receptive field} that depends on convolutional layer parameters such as the \emph{kernel\_size} and the \emph{dilation rate}:

\begin{equation}
     receptive field = 1+ \sum_{l=1}^L \ [k(l) - 1]\times d(l)
     \label{eq:rec_field}
\end{equation}

where $l \in {1,2 ... L}$ is a layer of the network.
This can be thought of as a sliding processing window for the input sequence.

\begin{figure}[h]
    \centering
    \includegraphics[width = \columnwidth]{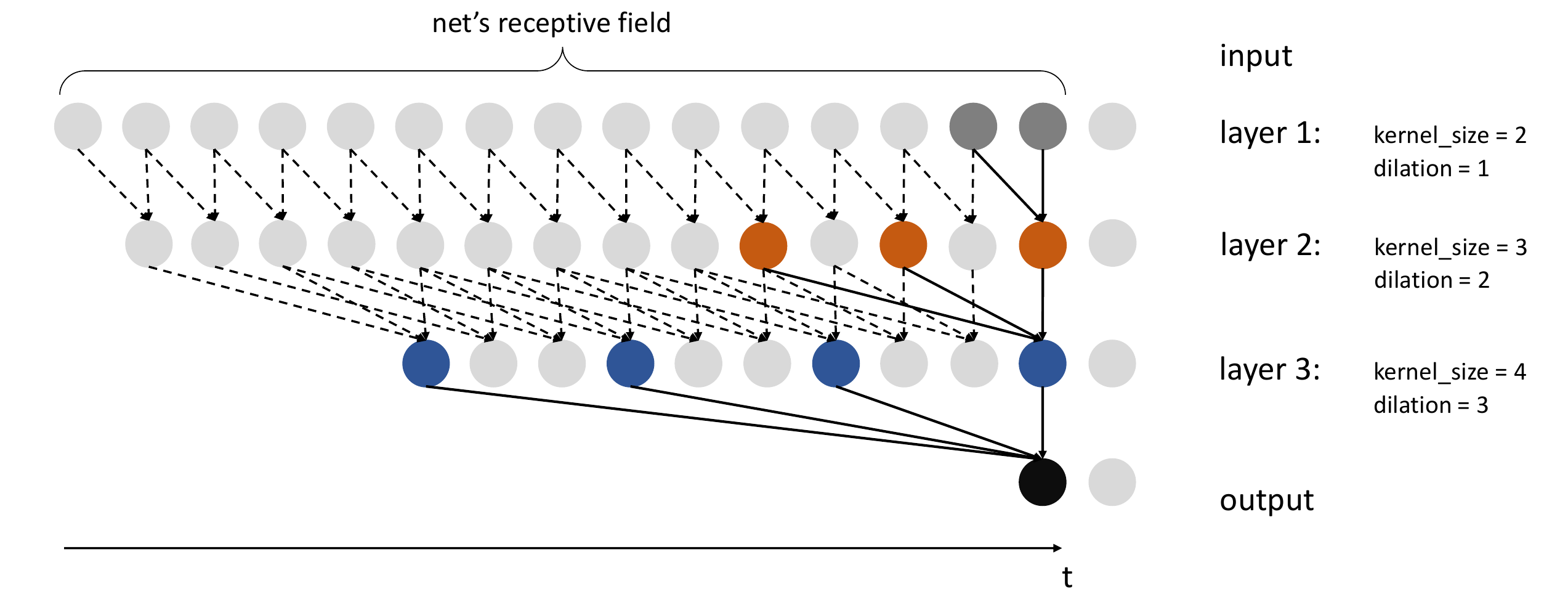}
    \caption{TCN execution graph. For this example, see (\ref{eq:rec_field}), $receptivefield=1+(2-1)\times 1+(3-1)\times 2+(4-1)\times3=15$}
    \label{fig:tcn_rec_field}
\end{figure}
Figure \ref{fig:tcn_rec_field} highlights these concepts.
It is worth noticing that increasing the dilation parameter leads to an increase of the network's memory without affecting its depth.
%As shown by equation \ref{eq:rec_field_wrkd_out} referred to the example network of Figure \ref{fig:tcn_rec_field}.

%\begin{equation}
%        rec field= 1+ (2-1)\times 1 + (3-1)\times 2 + (4-1)\times 3 = 15
%     \label{eq:rec_field_wrkd_out}
%\end{equation}

\section{Reference Accelerator Architecture}
\label{sec:ref_arch}
\begin{figure}[h]
    \centering
    \includegraphics[width = \columnwidth]{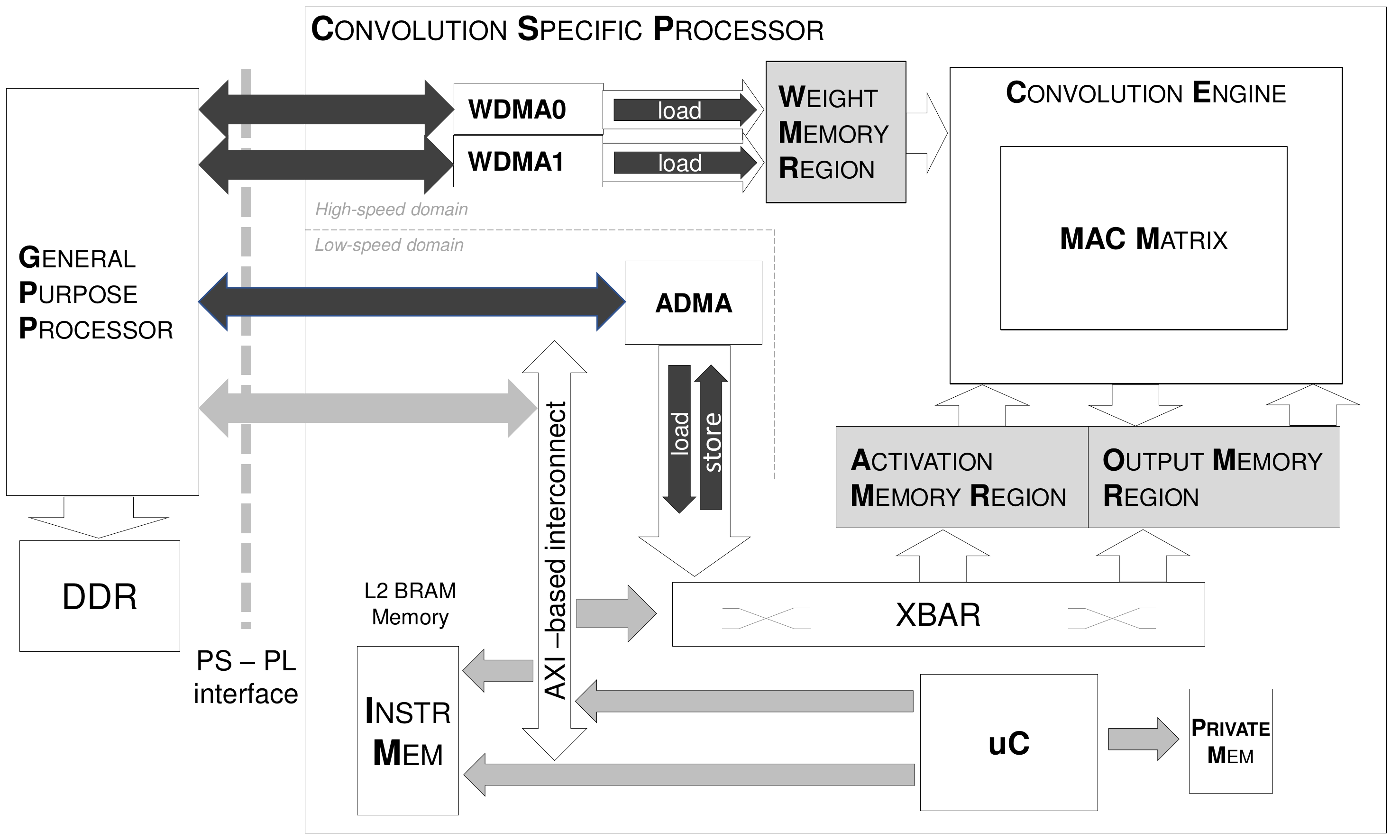}
    \caption{NEURAghe architecture}
    \label{fig:neu_arch}
\end{figure}
\begin{figure}[h]
    \centering
    \includegraphics[width = \columnwidth]{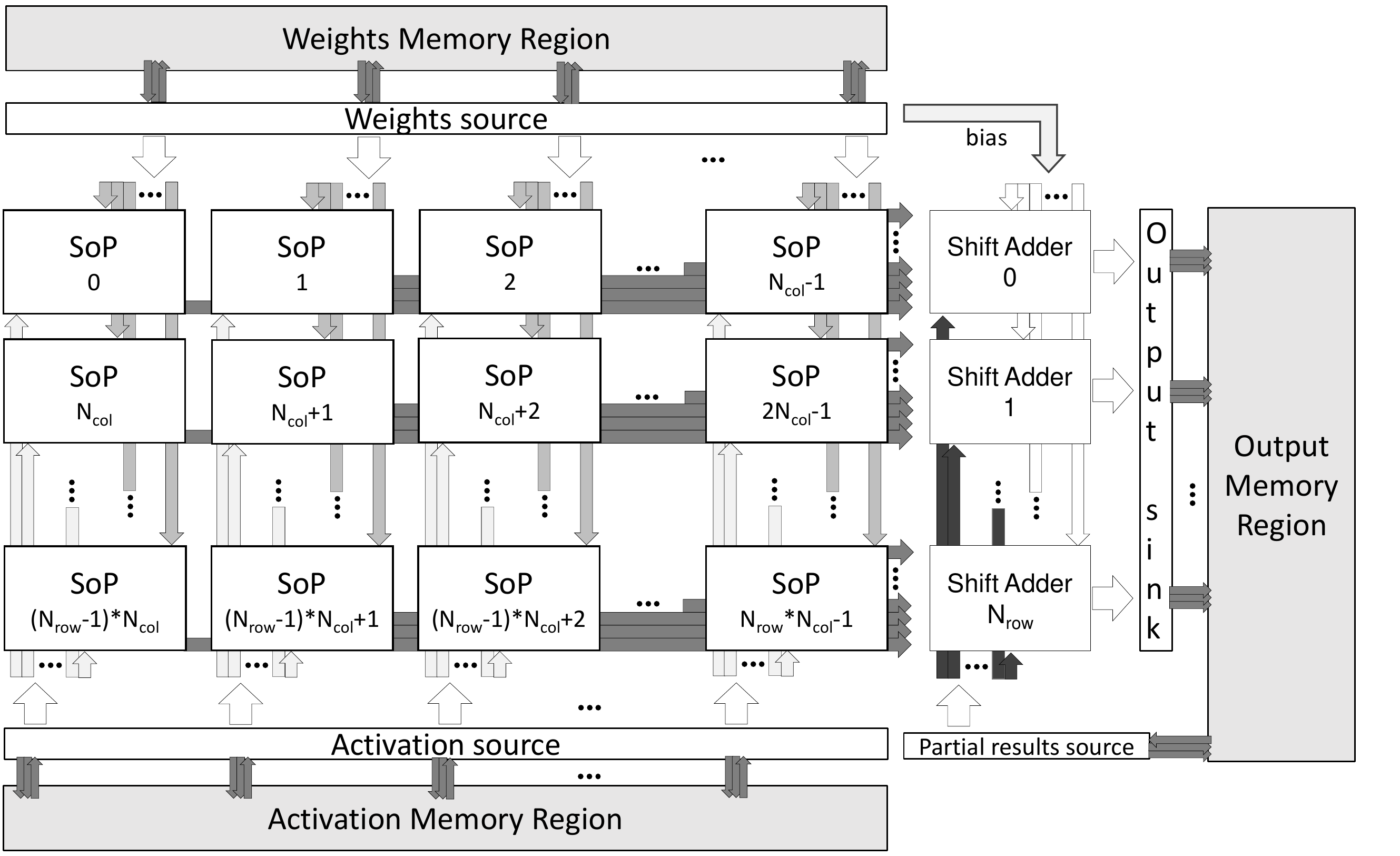}
    \caption{Convolution Engine. $N_{rows}\times N_{cols}$ MAC Matrix}
    \label{fig:ce}
\end{figure}

Our TCN accelerator is an extension of NEURAghe \cite{neu2017}, a CNN inference accelerator architecture that exploits the cooperation between the ARM Cortex-A9 Processing System (PS) and the Programmable Logic (PL) in Xilinx Zynq devices. Communication at the PS-PL interface is allowed by the PS high-performance general purpose ports. In NEURAghe, as can be seen in Figure \ref{fig:neu_arch}, the Programmable Logic hosts a Convolution Specific Processor (CSP), while the processing system acts as a General Purpose Processor (GPP) dealing with tasks hard to accelerate by parallelization on programmable logic, such as data marshalling or non-Conv layers.
NEURAghe integrates a RISC-V lightweight processor inside the CSP, dedicated to the execution of a firmware that schedules data transfers and convolutions without PS intervention, leaving the latter available for actual computation workload.
In order to support TCN execution, we have enhanced the CPS IP, described in System Verilog and available as open source \footnote{\urlstyle{tt}\url{https://github.com/neuraghe/NEURAghe}}. We have added new features to improve flexibility, requiring substantial modification of the main computational core dedicated to Multiply and Accumulate (MAC) operations execution, called Convolution Engine, and an improvement of the circuitry and procedures managing transfers to/from the DDR memory. %We describe in Sections \ref{CE} and \ref{SCHED} the TCN-supporting implementations.
In particular, the Convolution Engine (Figure \ref{fig:ce}) is composed by a MAC Matrix of $N_{cols}$ columns by $N_{rows}$ rows of Sum of Product (SoP) units in charge to calculate the contribution of $N_{cols}$ input features to $N_{rows}$ output features. $N_{rows}$ Shift Adder modules sum together partial result from SoPs in each row with data values resulted from the $N_{cols}$ previously computed input feature partial results read from on-chip memory, enabling successive accumulation over multiple CE runs.

%\subsection{Convolution Engine characteristics}\label{CE}
\section{TCN supporting hardware features}\label{sec:CE}

%The Convolution Engine in NEURAghe is devoted to execute the most computational-intensive tasks of a CNN, Multiply and Accumulate (MAC) operations, in parallel.
%It is composed by a matrix of $M$ columns by $N$ rows of Sum of Product (SoP) units in charge to calculate the contribution of $M$ input features to $N$ output features. Partial results from SoPs in each row are summed together by means of $N$ Shift Adder modules.

%Figure \ref{fig:ce} shows a specific SoP Matrix configuration implemented on a Zynq Z7020 All-Programmable SoCs.

%Shift Adder modules read data values resulted from the $M$ previously computed IF partial results from on-chip memory, to enable successive accumulation over multiple CE runs. 

Considering the typical features of TCNs, we designed the CE according to several design principles. The accelerator:
\begin{itemize}
    \item has to be $kernel\_size$ agnostic; 
    \item must execute convolutions with multiple stride values without performance overhead;
    \item must support freely selectable dilation values.
\end{itemize}

\subsection{Freely selectable kernel sizes}

\begin{figure}[h]
    \centering
    \includegraphics[width = \columnwidth]{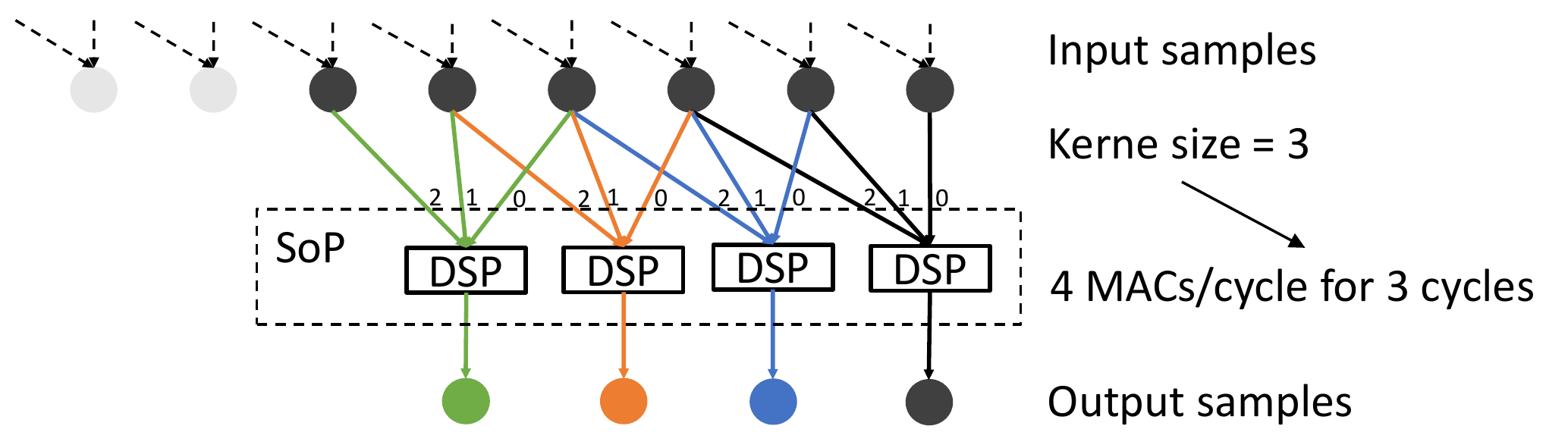}
    \caption{SoP elaboration scheme}
    \label{fig:sop_elab}
\end{figure}

To support arbitrary kernel sizes, we have chosen to dedicate one single DSP cell to compute an entire convolution kernel, reusing it over a number of cycles depending on the kernel size. A new sample of the output feature under production is thus produced after $kernel\_size$ cycles and is ready to be sent to the Shift Adder module. Using one single DSP cell per kernel can easily require the instantiation of a very high number of SoPs, to the aim of exploiting as many resources as possible among those available on the target device.
To keep the MAC Matrix growth feasible, we designed SoPs to be composed by $4$ Xilinx DSP48E primitives, performing $4\ MACs/cycle$, operating in parallel on $4$ different 16-bit samples, as visualized in Figure \ref{fig:sop_elab}, from $4$ neighbour convolution windows in an input feature.

\begin{figure}[h]
    \centering
    \includegraphics[width = \columnwidth]{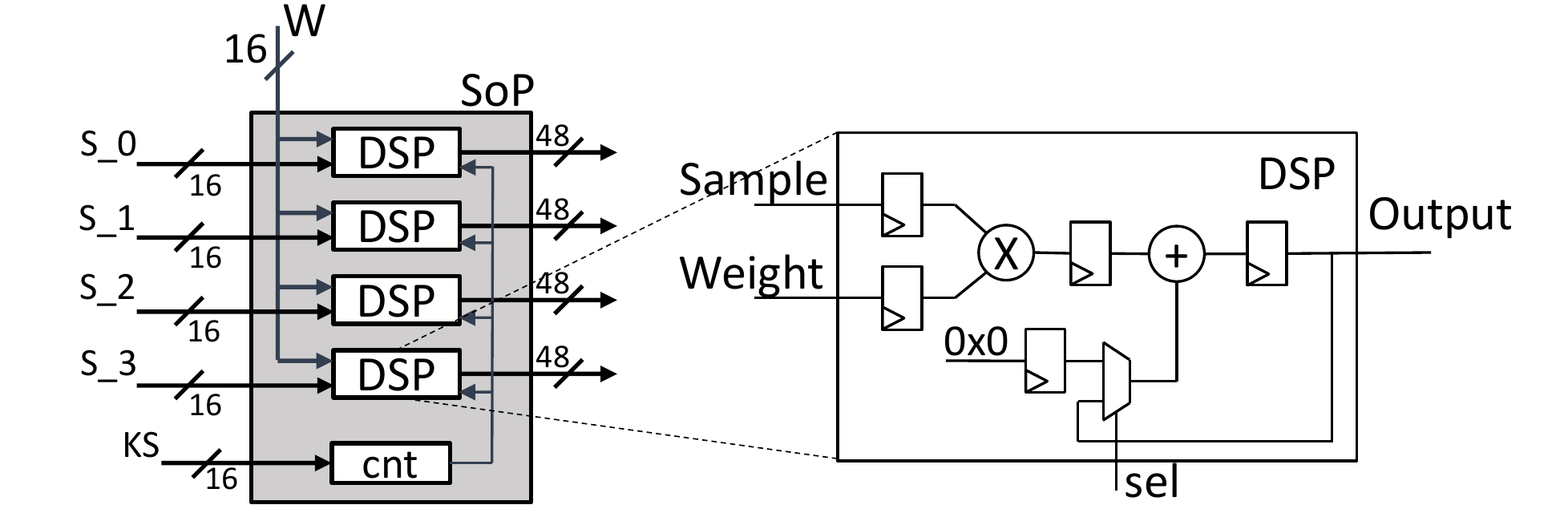}
    \caption{Sum of Product Unit}
    \label{fig:sop}
\end{figure}

Figure \ref{fig:sop} represents the organization of a SoP.
Considering the template proposed in Figure \ref{fig:ce}, the MAC Matrix implementation requires a deterministic number of DSP48E primitives, as indicated by Equation \ref{DSPscount}.
\begin{equation}\label{DSPscount}
    N_{DSPs}= N_{rows}\times N_{cols}\times 4
\end{equation}

\subsection{Flexible activations and weights fetching}

\begin{figure*}[]
    \centering
    \includegraphics[width = \textwidth]{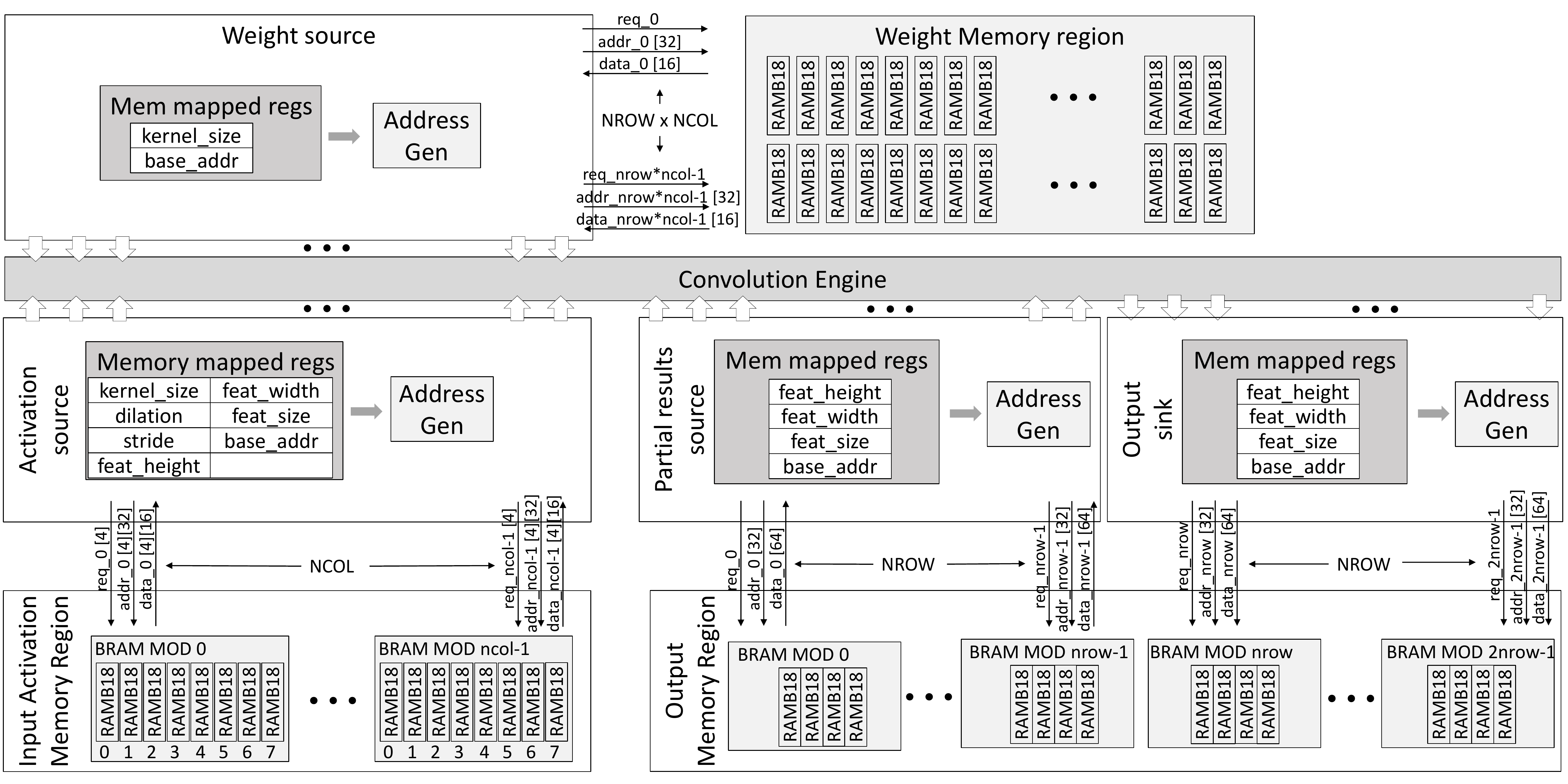}
    \caption{Memory transfers}
    \label{fig:act_fetch}
\end{figure*}

To enable arbitrary stride and dilation values, the fetching of input samples from the internal memory has been designed to be very flexible. Each CE port dedicated to input samples is endowed by a programmable Activation Source module, while weight kernels are fetched from Weight Memory banks by means of Weights Source modules. Such source modules can be programmed at start-up according to stride, dilation, aspect ratio and size of activations and weight kernels. Fetching of bi-dimensional memory sections is enabled, to support generic CNN execution. \par
\begin{figure}[H]
    \centering
    \includegraphics[width = \columnwidth]{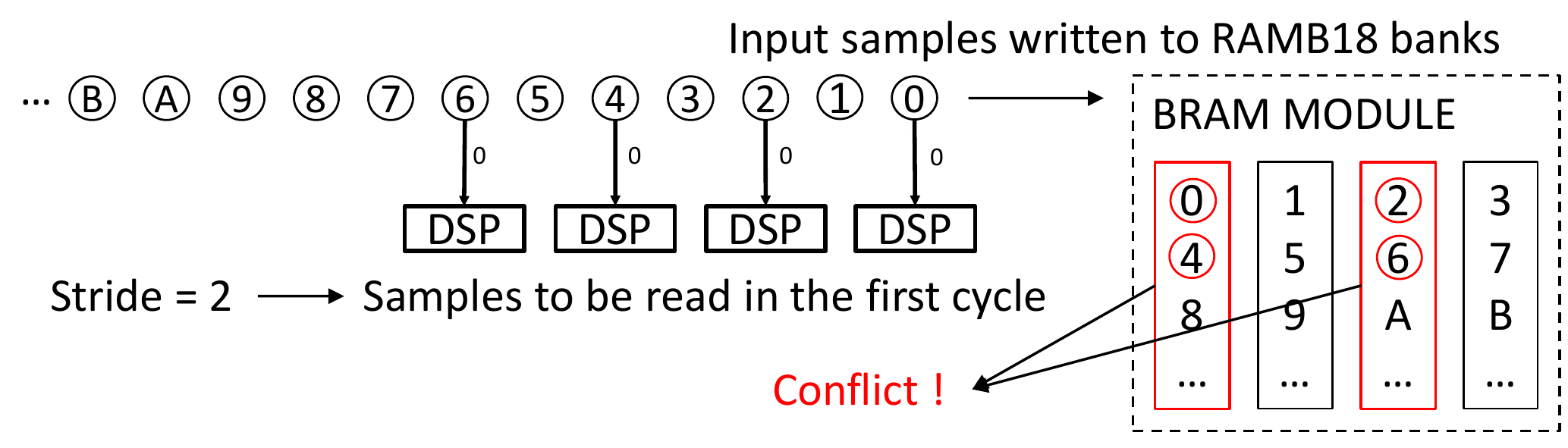}
    \caption{BRAM read conflict example. Samples are stored in BRAM modules using interleaving. Consecutive samples are stored in adjacent RAMB18 banks. 4 DSP slices in a SoP units, with stride 2, in the first cycle of the convolution, load respectively samples 0,2,4,6. With four RAMB18 banks, samples 0-4 and 2-6 are on the same bank, creating a conflict.}
    \label{fig:conflict_bram}
\end{figure}

\noindent
The memory subsystem has been designed to enable conflict-less loading of neighbour convolution windows. The Activation Source module controls $N_{cols}$ ports of the convolution engine. Each one loads samples from a dedicated BRAM module. Considering that four samples, belonging to different windows, can be loaded in the same cycle, to support stride values up to 3 samples (which is sufficient to support most of the TCN use-cases available in literature), each module of the activation memory has to be composed of at least 8 different independently accessible RAMB18 modules. Figure \ref{fig:conflict_bram} represents an example where a different configuration of RAMB18 modules can determine a conflict.\par\noindent

Moreover, the Weight source module controls one port per each SoP in the matrix, which has to be implemented by at least one RAMB18 module. \par\noindent
Finally, the CE has a set of ports that are used to write results and to load previously computed partial results, when a convolution requires to accumulate over several accelerator operations. These ports are controlled by a Partial Result Source module and by an Output Sink module. Each of these modules controls $N_{rows}$ ports, each one writing/reading four samples simultaneously. Thus BRAM modules in the corresponding memory region are composed by at least 8 RAMB18 modules each. \par\noindent Samples and weights, in the experiments presented in this paper, are all using a 16-bit data format, thus RAMB18 modules are configured to expose two 16-bit addressable ports and can be 1024 words deep. \par\noindent
Considering the described organization, a given architectural configuration requires a number of RAMB18 primitives that can be deterministically estimated as indicated in Equation \ref{BRAMScount}. 
\begin{equation}\label{BRAMScount}
    N_{BRAMs}= N_{rows}\times N_{cols} \\
    + N_{cols} \times 8 \\
    + (N_{rows}\times 2)\times 8 + \\
    32
\end{equation}
The first component corresponds to weight memory, the second to the modules storing input activations, the third to output and partial results memories. 32 blocks are used to implement the RISC-V scheduler instruction memory and private memory.

%This design allows the accelerator:
%\begin{itemize}
%    \item to be $kernel \ size$ agnostic; 
%    \item to execute convolutions with multiple stride values without performance overhead.
%\end{itemize}

%Shift Adder modules read 64-bit data that are $4$ 16-bit values resulted from the previous $M$ IF contribution and produce 64-bit data output summing together these inputs and those given by the actual IF contribution.

\subsection{TCN support in firmware}\label{sec:SCHED}

\begin{figure}[H]
    \centering
    \includegraphics[width = \columnwidth]{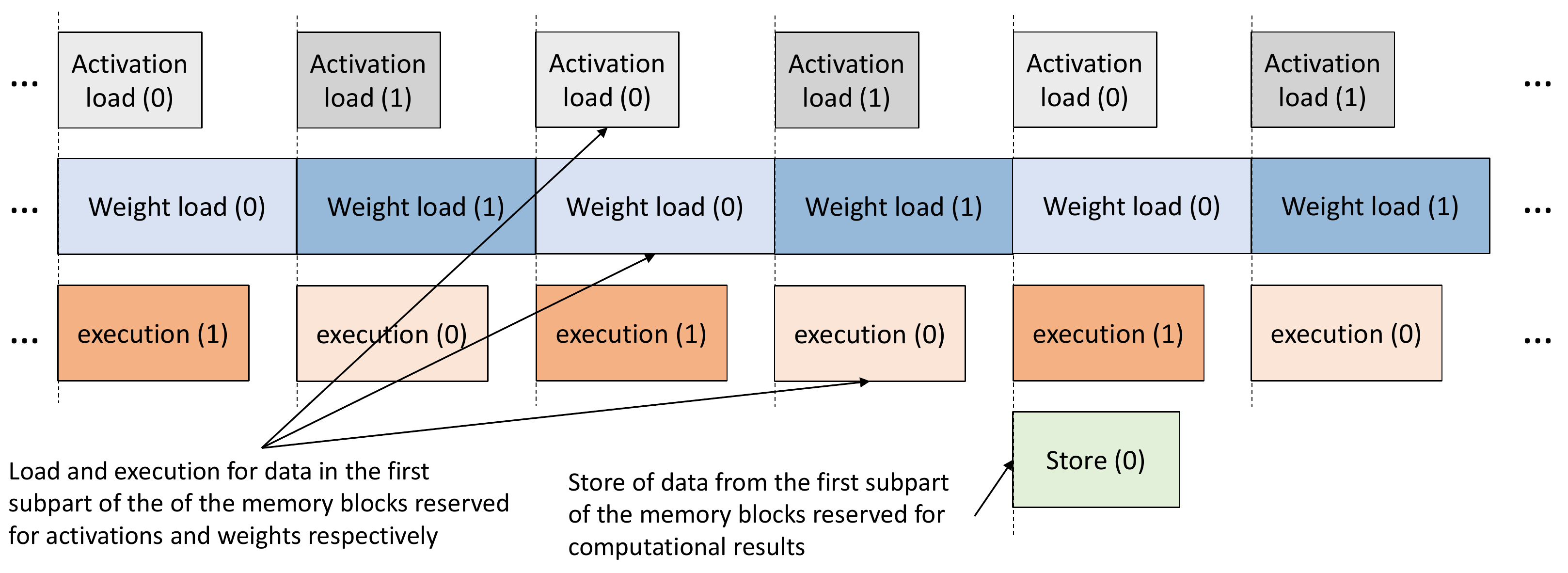}
    \caption{Scheduling scheme}
    \label{fig:scheduling}
\end{figure}

The programming model used in NEURAghe envisions the ARM-based processing system in the Zynq platforms to execute a C program implementing the neural network inference. When the accelerator implemented in the programmable logic has to be used. The program in the PS sends commands describing the layer to be executed. The RISC-V soft-core in the accelerator decodes the command and decomposes the layer in sub-operations, namely partial convolutions in the accelerator and data transfers from/to the off-chip memory, executing an optimized firmware which is also coded in C. The firmware uses a double-buffering technique, to allow the accelerator to overlap transfers phases with convolutions, as represented in Figure \ref{fig:scheduling}, reducing as much as possible idle times in the CE to exploit the processing capabilities of DSP slices with maximum efficiency.

%\subsection{TCN handling}

% \begin{figure}[]
%     \centering
%     \includegraphics[width = \columnwidth]{img/hwce_feed.pdf}
%     \caption{TCN execution on NEURAghe}
%     \label{fig:tcn_exec}
% \end{figure}

When considering TCN executions, the previously described paradigm has to be applied to the characteristics of the algorithm. For every Convolutional Layer in a TCN, given its \emph{kernel\_size} and a \emph{dilation}, it is possible to consider a local \emph{receptive field}, indicated as $RF_{local}$ in Equation \ref{loc_rec_field}, that is the minimum amount of layer's input samples per channel needed to produce a valuable output sample.

\begin{equation}\label{loc_rec_field}
    RF_{local}= 1 + (k - 1)\times d
\end{equation}

\begin{figure}[h]
    \subfloat[\label{fig:B=1}]{\includegraphics[width = 0.5\columnwidth]{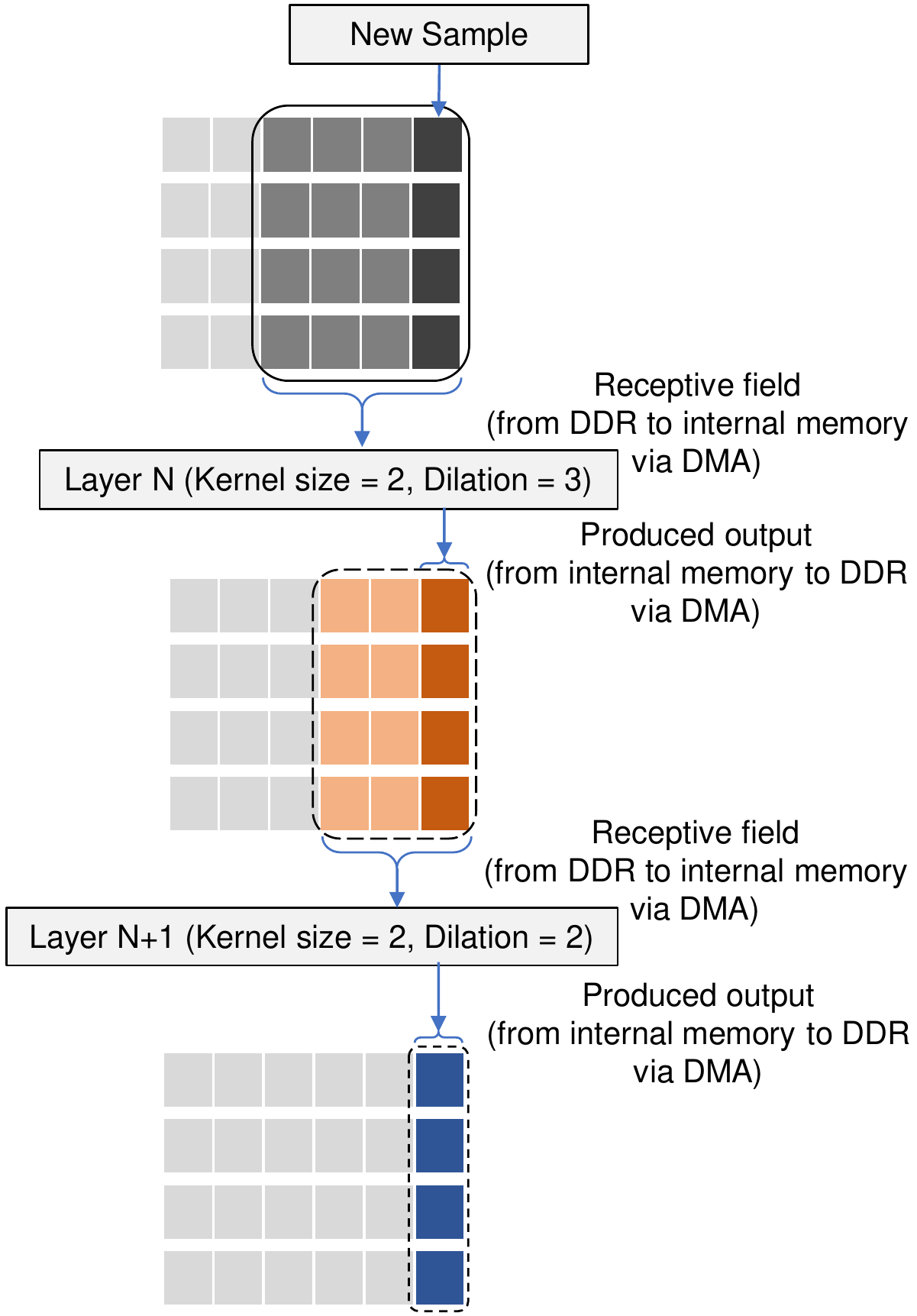}}
    \hfill
    \subfloat[\label{fig:B=B}]{\includegraphics[width = 0.5\columnwidth]{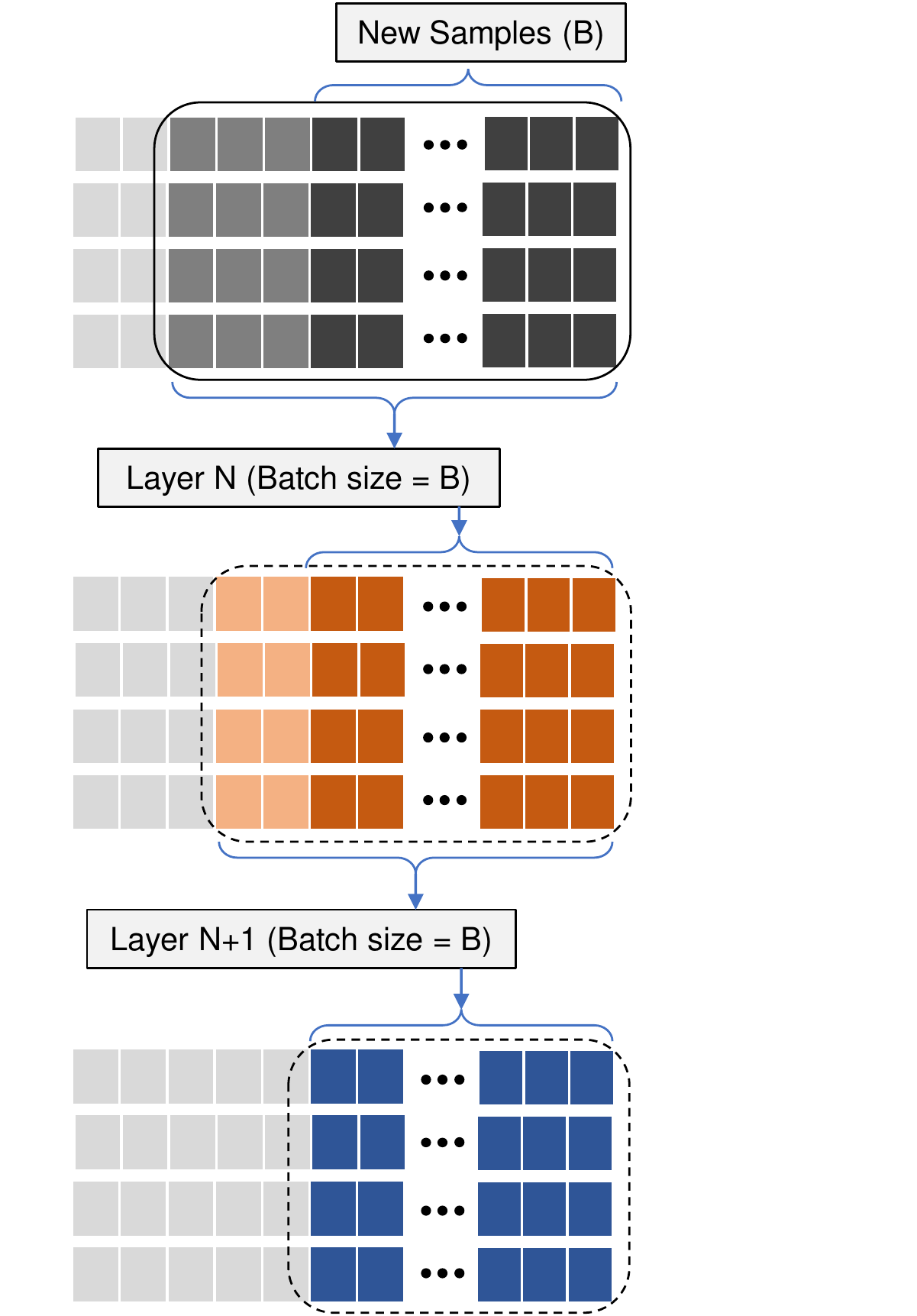}}
    \caption{TCN execution on NEURAghe}
    \label{fig:tcn_exec}
\end{figure}

Figure \ref{fig:B=1} shows how input and output transfers are implemented for TCNs.
At every new time step, input to a layer is updated by adding one new sample to input features. In order to execute the layer, the firmware triggers DMA transfers to load to the activation input memory $RF_{local}$ samples per each input feature. After the execution, one output sample is produced per output feature. Output samples are sent to DDR using an output DMA transfer, to be stored until the next time step. 

\subsection{Improving through batch processing} \label{batch proc}

Although the previous approach provides the minimum classification/recognition latency, executing the network every time a new sample is available to update the input sliding window, it can determine performance to be bandwidth limited. 
This is because all of network parameters/weights must be loaded for every layer. So, despite the double-buffered scheduling strategy, transfer and computation phases hardly overlap, affecting the \emph{operational intensity} of the application.
\begin{figure}[h]
    \centering
    \includegraphics[width = \columnwidth]{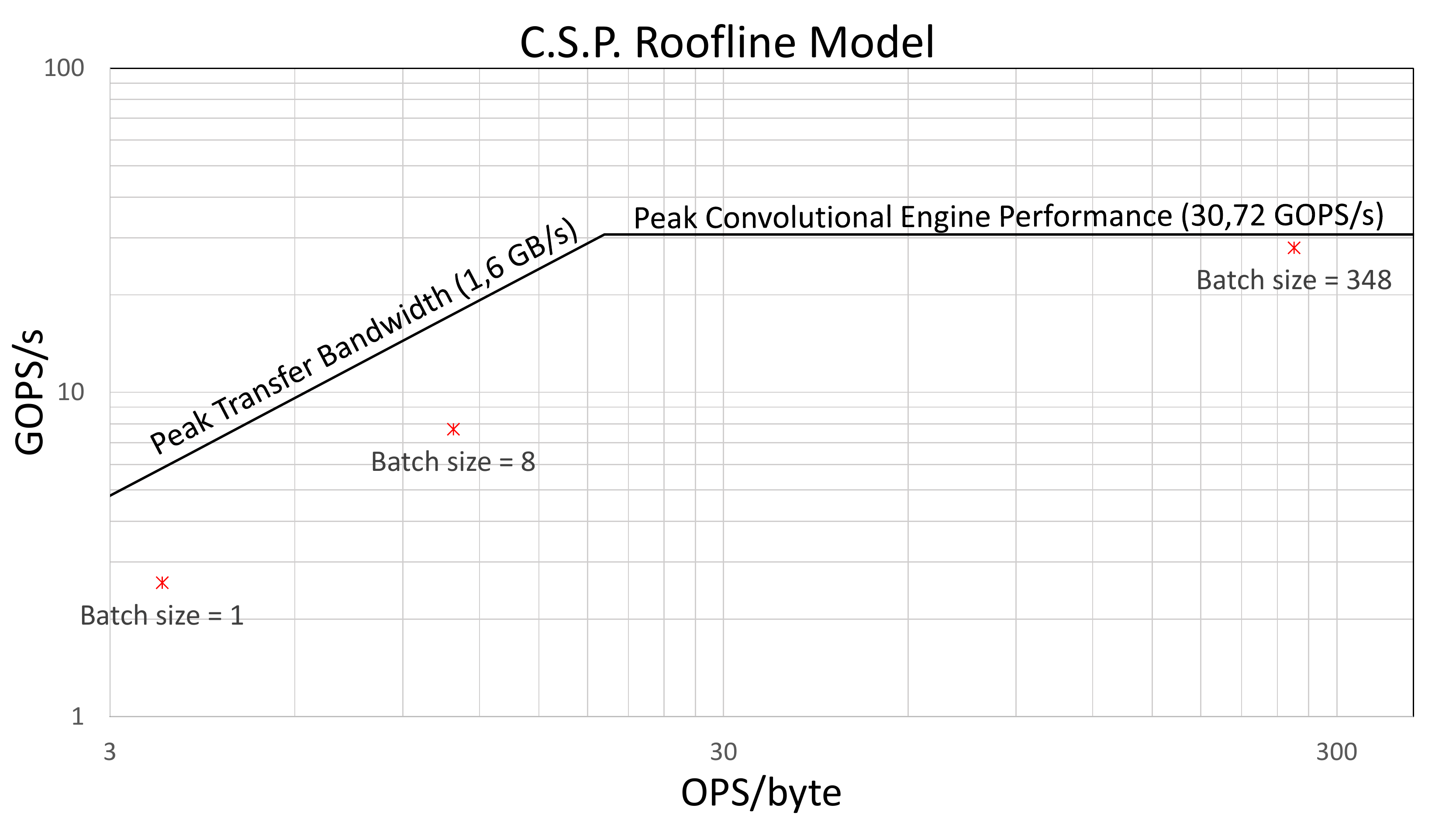}
    \caption{Use case benchmark's Roofline Model for the Convolution Specific Processor (12x4 MAC Matrix in Z-7020 SoC) with respect to different batch sizes}
    \label{fig:roofline}
\end{figure}
The roofline model in Figure \ref{fig:roofline} shows performance trend starting from the sample-by-sample processing (leftmost red cross), on a use-case that will be presented in the following.

% \begin{figure}[]
%     \centering
%     \includegraphics[width = \columnwidth]{img/hwce_feed_rec_field_batched_3.pdf}
%     \caption{TCN execution with batching on NEURAghe}
%     \label{fig:tcn_model_batched_on_neuraghe}
% \end{figure}

If the latency constraint is not extremely tight, it is possible to pre-buffer input samples in order to process longer sample sequences (batch) and produce more output with every execution. 
Figure \ref{fig:B=B} shows the transfer scheduling when \emph{batch size} is increased.
As the \emph{batch size} increases the architecture gets closer and closer to the computational limit (rightmost red cross on the roofline plot of Figure \ref{fig:roofline}) because of the growing number of operations performed. In this way, it is possible to increase the utilization of computing resources and to gain efficiency.
As may be noticed in Figure \ref{fig:roofline}, when sample batch size is small, besides being on the badwidth-limited region of the plot, points in the roofline model are also distant from the theoretical achievable performance. This is due to overheads related with CE programming/warm-up,  and to initial and final input/output transfers, which cannot overlap with convolutions. The impact of this overhead is limited when the operational intensity increases, reducing the distance between points and the theoretical roofline model.   

%\subsection{Resources utilization on target board}
%resource utilization model that links NROWs, NCOLs and with

\section{Hardware Implementation Evaluation}
\label{sec:hw_impl_eval}

\subsection{Design Space Exploration}

A designer willing to use the NEURAghe template, on a given target SoC, has multiple possible architectural configurations available. In order to perform a careful selection, it is possible to perform a simple design space exploration and to choose a near-optimal setup from the performance point of view. To select the architectures presented in this paper, we have used a simple grid search that evaluates multiple configurations, featuring different values of $N_{rows}$ and $N_{cols}$, to maximize the number of SoPs, while keeping the number of used DSPs and BRAMs in the range of availability imposed by the target SoC. 
Table \ref{tab:board_utiliz} shows utilization numbers estimated using the Equations \ref{DSPscount} and \ref{BRAMScount}. Light-gray coloured cells in the table indicate which configurations are not implementable in a Z7020 SoC, due to excessive DSP or BRAM utilization. Darker-coloured cells indicate configurations that are not feasible in an Ultrascale+ ZU3EG device. 

\begin{table}[h]
    \caption{RAMB18 and DSP utilization in NEURAghe architecture with respect to the MAC Matrix shape}
    \begin{center}
    \setlength{\arrayrulewidth}{0.3mm}
    \renewcommand{\arraystretch}{1.2}
    \begin{tabular}{*{1}{|c}*{1}{|c}|*{9}{c|}}
                \cline{1-11}
                    &   \multicolumn{9}{c|}{$N_{cols}$}  & \\
                \cline{1-11}
                                            &    \tikzmark{startbram} 144 \tikzmark{endbram}    &    156    &    168    &    180    &    192    &    204    &    216    &    228    &    240    &    \multirow{2}{0.8em}{4}\\
                                            &    \tikzmark{startdsp} 64 \tikzmark{enddsp}    &    80    &    96    &    112    &    128    &    144    &    160    &    176    &    \tikzmark{startone} 192 \tikzmark{endone}    &    \\
                \cline{2-11}
                                            &    164    &    177    &    190    &    203    &    216    &    229    &    242    &    255    &    268    &    \multirow{2}{0.8em}{5}\\
                                            &    80    &    100    &    120    &    140    &    160    &    180    &    200    &    \tikzmark{starttwo} 220 \tikzmark{endtwo}    &    \cellcolor{lightgray}240    &    \\
                \cline{2-11}
                                            &    184    &    198    &    212    &    226    &    240    &    254    &    268    &    \cellcolor{lightgray}282    &    \cellcolor{lightgray}296    &    \multirow{2}{0.8em}{6}\\
\multirow{9}{0.8em}{\rotatebox{90}{$N_{rows}$}}    &    96    &    120    &    144    &    168    &    192    &    216    &    \cellcolor{lightgray}240    &    \cellcolor{lightgray}264    &    \cellcolor{lightgray}288    &    \\
                \cline{2-11}
                                            &    204    &    219    &    234    &    249    &    264    &    279    &    \cellcolor{lightgray}294    &    \cellcolor{lightgray}309    &    \cellcolor{lightgray}324    &    \multirow{2}{0.8em}{7}\\
                                            &    112    &    140    &    168    &    196    &    \cellcolor{lightgray}224    &    \cellcolor{lightgray}252    &    \cellcolor{lightgray}280    &    \cellcolor{lightgray}308    &    \cellcolor{lightgray}336    &    \\
                \cline{2-11}
                                            &    224    &    240    &    256    &    272    &    \cellcolor{lightgray}288    &    \cellcolor{lightgray}304    &    \cellcolor{lightgray}320    &    \cellcolor{lightgray}336    &    \cellcolor{lightgray}352    &    \multirow{2}{0.8em}{8}\\
                                            &    128    &    160    &    192    &    \cellcolor{lightgray}224    &    \cellcolor{lightgray}256    &    \cellcolor{lightgray}288    &    \cellcolor{lightgray}320    &    \cellcolor{lightgray}352    &    \cellcolor{gray}384    &    \\
                \cline{2-11}
                                            &    244    &    261    &    278    &    \cellcolor{lightgray}295    &    \cellcolor{lightgray}312    &    \cellcolor{lightgray}329    &    \cellcolor{lightgray}346    &    \cellcolor{lightgray}363    &    \cellcolor{lightgray}380    &    \multirow{2}{0.8em}{9}\\
                                            &    144    &    180    &    216    &    \cellcolor{lightgray}252    &    \cellcolor{lightgray}288    &    \cellcolor{lightgray}324    &    \tikzmark{startthree} \cellcolor{lightgray}360 \tikzmark{endthree}    &    \cellcolor{gray}396    &    \cellcolor{gray}432    &    \\
                \cline{2-11}
                                            &    264    &    \cellcolor{lightgray}282    &    \cellcolor{lightgray}300    &    \cellcolor{lightgray}318    &    \cellcolor{lightgray}336    &    \cellcolor{lightgray}354    &    \cellcolor{lightgray}372    &    \cellcolor{lightgray}390    &    \cellcolor{lightgray}408    &    \multirow{2}{0.8em}{10}\\
                                            &    160    &    200    &    \cellcolor{lightgray}240    &    \cellcolor{lightgray}280    &    \cellcolor{lightgray}320    &    \cellcolor{lightgray}360    &    \cellcolor{gray}400    &    \cellcolor{gray}440    &    \cellcolor{gray}480    &    \\
                \cline{2-11}
                                            &    \cellcolor{lightgray}284    &    \cellcolor{lightgray}303    &    \cellcolor{lightgray}322    &    \cellcolor{lightgray}341    &    \cellcolor{lightgray}360    &    \cellcolor{lightgray}379    &    \cellcolor{lightgray}398    &    \cellcolor{lightgray}417    &    \cellcolor{gray}436    &    \multirow{2}{0.8em}{11}\\
                                            &    176    &    220    &    \cellcolor{lightgray}264    &    \cellcolor{lightgray}308    &    \cellcolor{lightgray}352    &    \cellcolor{gray}396    &    \cellcolor{gray}440    &    \cellcolor{gray}484    &    \cellcolor{gray}528    &    \\
                \cline{2-11}
                                            &    \cellcolor{lightgray}304    &    \cellcolor{lightgray}324    &    \cellcolor{lightgray}344    &    \cellcolor{lightgray}364    &    \cellcolor{lightgray}384    &    \cellcolor{lightgray}404    &    \cellcolor{lightgray}424    &    \cellcolor{gray}444    &    \cellcolor{gray}464    &    \multirow{2}{0.8em}{12}\\
                                            &    192    &    \cellcolor{lightgray}240    &    \cellcolor{lightgray}288    &    \cellcolor{lightgray}336    &    \cellcolor{gray}384    &    \cellcolor{gray}432    &    \cellcolor{gray}480    &    \cellcolor{gray}528    &    \cellcolor{gray}576    &    \\
         \hline
                                            &    4        &    5        &    6        &        7    &    8        &    9        &    10    &    11    &    12    & \\
         \hline
    \end{tabular}
    \end{center}
    \label{tab:board_utiliz}
    \begin{center}
    \begin{tabular}{*{1}{c}*{4}{|c}}
                &   avail.  &   avail. &                                        &   out of        \\
                &   RAMB18  &   DSP &   selected                                &   resources        \\
\hline
         Z7020  &   280     &   220 &   \tikzmark{startlone}     \tikzmark{endlone}    &   \cellcolor{lightgray}   \\
\hline
         ZU3EG  &   432     &   360 &   \tikzmark{startltwo}      \tikzmark{endltwo}    &   \cellcolor{gray}        \\
    \end{tabular}
    \end{center}
    
\begin{tikzpicture}[remember picture,overlay]

\draw[rounded corners,black,thick]
  ([shift={(-0.5\tabcolsep,-0.5ex)}]pic cs:startbram) 
    rectangle 
  ([shift={(0.5\tabcolsep,2ex)}]pic cs:endbram);
  
\draw [ -stealth ]
    ([shift={(0\tabcolsep,0)}]pic cs:startbram) 
        --
    ([shift={(-11\tabcolsep,0.1)}]pic cs:startbram) node [above] {RAMB};

\draw[rounded corners,black,thick]
  ([shift={(-0.5\tabcolsep,-0.5ex)}]pic cs:startdsp) 
    rectangle 
  ([shift={(0.5\tabcolsep,2ex)}]pic cs:enddsp);
  
\draw [ -stealth ] 
    ([shift={(0\tabcolsep,0)}]pic cs:startdsp) 
        --
    ([shift={(-8\tabcolsep,-0.1)}]pic cs:startdsp) node [left] {DSP};

\foreach \Val in {one,two}
{
\draw[rounded corners,green,thick]
  ([shift={(-0.5\tabcolsep,-0.5ex)}]pic cs:start\Val) 
    rectangle 
  ([shift={(0.5\tabcolsep,4.8ex)}]pic cs:end\Val);
}

\draw[rounded corners,red,thick]
  ([shift={(-0.5\tabcolsep,-0.5ex)}]pic cs:startthree) 
    rectangle 
  ([shift={(0.5\tabcolsep,4.8ex)}]pic cs:endthree);

\draw[rounded corners,green,thick]
  ([shift={(-0.5\tabcolsep,-0.5ex)}]pic cs:startlone) 
    rectangle 
  ([shift={(1\tabcolsep,1ex)}]pic cs:endlone);

\draw[rounded corners,red,thick]
  ([shift={(-0.5\tabcolsep,-0.5ex)}]pic cs:startltwo) 
    rectangle 
  ([shift={(1\tabcolsep,1ex)}]pic cs:endltwo);
\end{tikzpicture}
\end{table}

%\begin{algorithm}[h]
%% \KwData{this text}
%% \KwResult{how to write algorithm with \LaTeX2e }
%% initialization\;
% setup baseline\; 
% \While{there are cores available}{
%  locate the bottleneck\;
%  apply the splitting mechanism on the bottleneck\;
% }
% \While{true}{
%  wait trigger from periodic timer\;
%  check workload\;
%  \eIf{real-time constraints are not respected}{
%   Increase system frequency\;
%   Increase supply voltage\;
%   }{
%    Decrease system frequency\;
%    Decrease supply voltage\;
%  }
% }
% \caption{ADAM-FF policy algorithm.}\label{alg:ADAM-FF}
%\end{algorithm}

\subsection{Implementation on different SoCs}
%
%The particular configuration of NEURAghe, with a Convolution Engine like the one shown in figure \ref{fig:ce}, has been implemented in a Zedboard which integrates a Xilinx Zynq Z-7020, clocked at 80 MHz.
%
We have implemented the NEURAghe configurations with different Convolution Engine's MAC Matrix shapes, selected after the DSE process, on two target platforms: a Xilinx Zynq Z-7020 and a Xilinx Zynq Ultrascale+ ZU3EG. For the Z-7020 we have selected two configurations, featuring a similar number of DSP slices and similar clock frequencies. Table \ref{tab:occup_12x4} shows resource occupation on the reconfigurable logic of a first configuration implemented on the Z-7020, featuring a 12$\times$4 MAC matrix of SoP units, that can be clocked up to 120 MHz, providing peak performance of 46 GOPS/s. Table \ref{tab:occup_11x5} shows results related with a similar configuration that integrates an 11$\times$5 MAC matrix, using slightly more DSPs but clockable up to 110 MHz, with a peak performance of 48.4 GOPS/s. Finally, Table \ref{tab:occup_ultra} describes the results of the implementation of a configuration, featuring a 9$\times$10 MAC matrix of SoP modules, on the ZU3EG. This design uses all the DSP slices on the chip and can be clocked at 180 MHz, providing a peak of 129.6 GOPS/s.
\begin{table}[h!]
    \caption{Resource occupation on a Xilinx Zynq Z-7020 (12x4 MAC matrix)}
    \centering
    \begin{tabular}{*{1}{c}||*{5}{c}}
                    &  DSP      & BRAM  & LUTs  & LUTs  & Regs \\
                    &           &       & (logic) & SR   &   \\
         \hline
         \hline
         Used       &  192   & 120    & 47230   & 259    & 26942 \\
         \hline
         Available  &  220    & 140   & 53200   & 53200    & 106400 \\
         \hline
         \%         &  87.27    & 85.71 & 88.78   & 1.49   & 25.32 \\
         \multicolumn{6}{c}{} \\
    \end{tabular}
    \label{tab:occup_12x4}
\end{table}
%
%The architecture exploits 192 out of the 220 DSP blocks available in the device, so the processing power utilization is very high.
%Also the Block RAM primitives are extensively used due to the particular Weights Memory implementation.
%
\begin{table}[h!]
    \caption{Resource occupation on a Xilinx Zynq Z-7020 (11x5 MAC matrix)}
    \centering
    \begin{tabular}{*{1}{c}||*{5}{c}}
                    &  DSP      & BRAM  & LUTs  & LUTs  & Regs \\
                    &           &       & (logic) & SR   &   \\
         \hline
         \hline
         Used       &  220   & 128    & 42964   & 283    & 26962 \\
         \hline
         Available  &  220    & 140   & 53200   & 53200    & 106400 \\
         \hline
         \%         &  100    & 91.4 & 80.76   & 1.63   & 25.34 \\
         \multicolumn{6}{c}{} \\
    \end{tabular}
    \label{tab:occup_11x5}
\end{table}
\begin{table}[h!]
    \caption{Resource occupation on a Xilinx Zynq UltraScale+ ZU3EG (9x10 MAC matrix)}
    \centering
    \begin{tabular}{*{1}{c}||*{5}{c}}
                    &  DSP      & BRAM  & LUTs  & LUTs  & Regs \\
                    &           &       & (logic) & SR   &   \\
         \hline
         \hline
         Used       &  360   & 354    & 47857   & 159    & 26463 \\
         \hline
         Available  &  360    & 432   & 70560   & 70560    & 141120 \\
         \hline
         \%         &  100    & 81.94 & 67.82   & 0.4   & 18.75 \\
         \multicolumn{6}{c}{} \\
    \end{tabular}
    \label{tab:occup_ultra}
\end{table}

%\section{Use case network: ECG Classification}
\section{Experimental Results}
\label{sec:exp_res}
We tested the actual level of performance that can be achieved using the accelerator on three different benchmarks, trying to cover as much as possible the landscape of TCNs presented in literature:
\begin{itemize}
    \item a \emph{plain} TCN network for ECG monitoring and classification (Goodfellow et al. \cite{ecg}), that performs classification over single lead ECG waveforms and reaches around $90\%$ average accuracy. We call this benchmark \emph{ECG} in the following. This benchmark exposes real-time constraints and can be used to assess the usability of the accelerator in this kind of context. 
    \item A network, called hereafter Res-TCN, based on residual units, with a structure taken from the Resnet family presented in \cite{resnet}. The TCN is presented in (\cite{3D_skeleton}) authors started from a 3D human action recognition dataset with 3D full skeleton annotations to extract a 1D feature representation per frame resulting in a 150-dimensional vector.
    \item  A more complex network, based on WaveNet  (Van  DenOord et al.  [14]),   for  Polyphonic Note  Transcription of Time-Domain Audio Signal ([19]). We indicate this use-case as WN-PNT.
\end{itemize}
We have considered three system implementations using NEURAghe, two of them, 11$\times$5 and 12$\times$4, implemented on a Zedboard development board, with the MAC matrix respectively clocked at 70 and 80 MHz, and one implemented on the Ultra96 development board, with MAC Matrix clocked at 180 MHz.

Table \ref{tab:mem_foot} shows the comparison between on-chip memory capabilities for each system implementation and the memory footprint of each use-case network that regards the overall occupation of weight kernels and input features through all layers, considering the sample batch size values that will be used in the following. As may be noticed, all the networks must be adequately managed with  the scheduling strategy presented in section \ref{sec:SCHED}, since their activation and weight memory footprint exceeds the memory available on the considered low-cost hardware platforms.

\begin{table}[h!]
    \caption{Implementation related on-chip memory capabilities and use-case networks memory footprint}
    \centering
    \begin{tabular}{*{1}{c}|*{3}{c}*{1}{c}*{1}{c}}

                    &   \multicolumn{3}{c}{input features [kB]}   & weight kernels [kB]   \\
         \hline
\multirow{2}{8em}{12x4 on Z-7020 on-chip memory}    & \multicolumn{3}{c}{\multirow{2}{1em}{192}}    &  \multirow{2}{1em}{96}   \\
            & & \\
         \hline
\multirow{2}{8em}{11x5 on Z-7020 on-chip memory}    & \multicolumn{3}{c}{\multirow{2}{1em}{176}}    &  \multirow{2}{1em}{110}   \\
            & & \\
         \hline
\multirow{2}{8em}{9x10 on ZU3EG on-chip memory}  & \multicolumn{3}{c}{\multirow{2}{1em}{144}}    &   \multirow{2}{1em}{180}  \\
            & & \\
         \hline
         ECG:    &  \textbf{B=1}  & \textbf{B=8} & \textbf{B=348}    &               \\
         memory footprint   &  216,8    &   250,8     &   1696,5    &     11495,6  \\
         \hline
         Res-TCN:    &  \textbf{B=1} &  \multicolumn{2}{c}{\textbf{B=144}}    &    \\
         memory footprint   &      37,3         &  \multicolumn{2}{c}{511,5}    &  5424,8  \\
         \hline
         WN-PTN:     &  \textbf{B=1}   &  \multicolumn{2}{c}{\textbf{B=504}}      &   \\
         memory footprint   &      553,5     &   \multicolumn{2}{c}{5846,49}  &   3328,8    \\
        %  \hline
        %  GOPS               &  0,044  &     0,132     &  3,858  &   \\
        %  GOPS               &  0,020         &  \multicolumn{2}{c}{0,684}   &   \\
        %  GOPS               &   0,007    &   0,675    &   1,705 &   \\
    \end{tabular}
    \label{tab:mem_foot}
\end{table}

\subsection{ECG classification use-case}
The network consists of several computational blocks mostly made of a 1D Convolutional layer, a batch normalization layer, a ReLU and a dropout stage, operating on streams of 16-bit samples, acquired at 300 Hz frequency.
In this case, we have assessed three different operating modes: 
\begin{itemize}
\item \textit{minimum latency mode}, 
\item \textit{maximum throughput mode}, 
\item \textit{real-time execution}.
\end{itemize}
The first one provides classification in output after minimum latency. The network is executed as soon as possible per every input sample. 
In the second operating mode, sample batching is used extensively, to maximize throughput without considering latency as an optimization objective. In the third mode, sample batching is exploited just enough to reach a throughput that allows respecting the real-time constraint posed by the input sampling frequency (300 Hz).

\begin{figure}[h]
    \centering
    \includegraphics[width = \columnwidth]{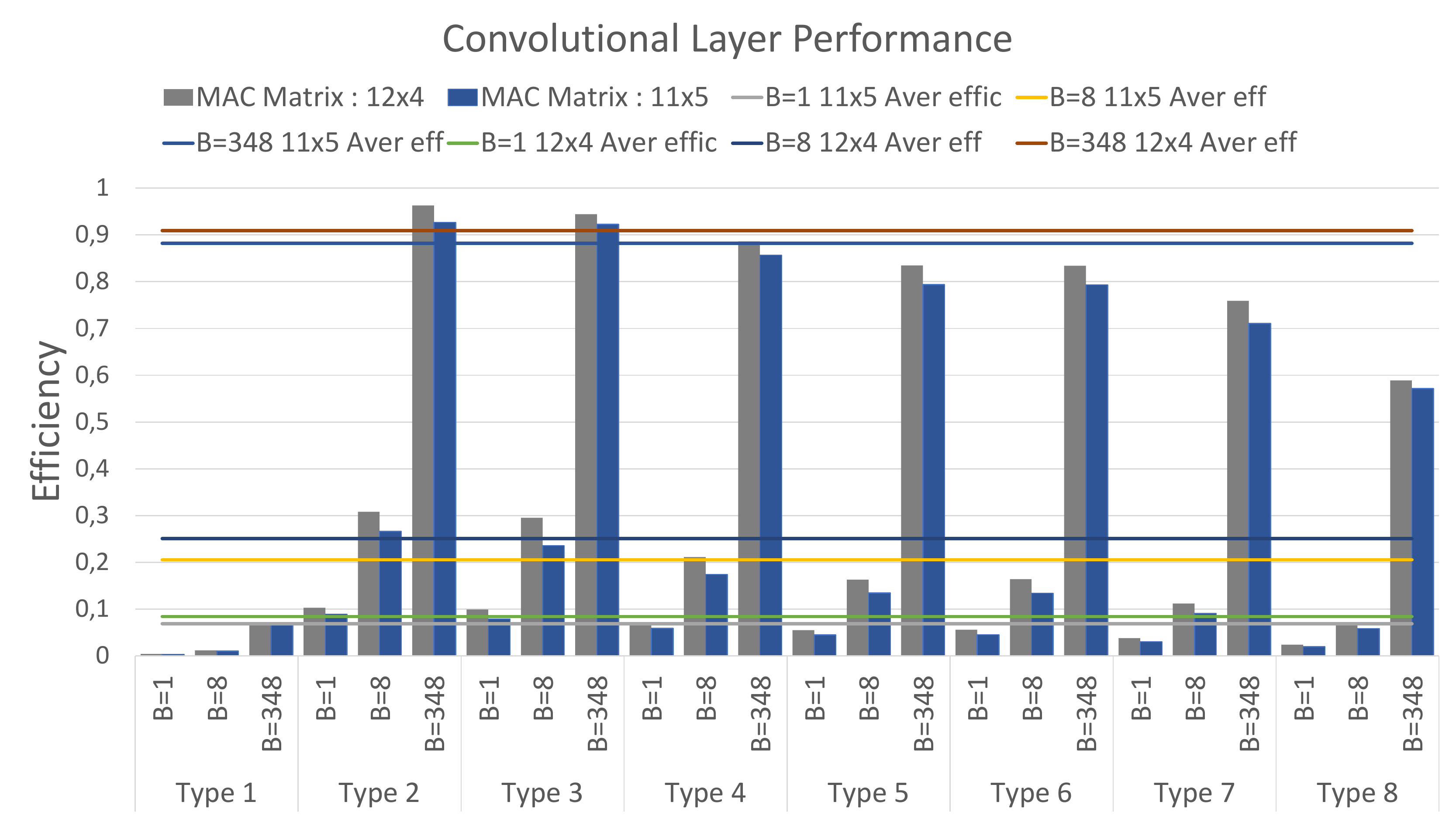}
    \caption{Efficiency comparison on ECG \cite{ecg} for NEURAghe 12x4 and 11x5 MAC matrix configurations}
    \label{fig:effic_compar_zed}
\end{figure}

\begin{figure}[h]
    \centering
    \includegraphics[width = \columnwidth]{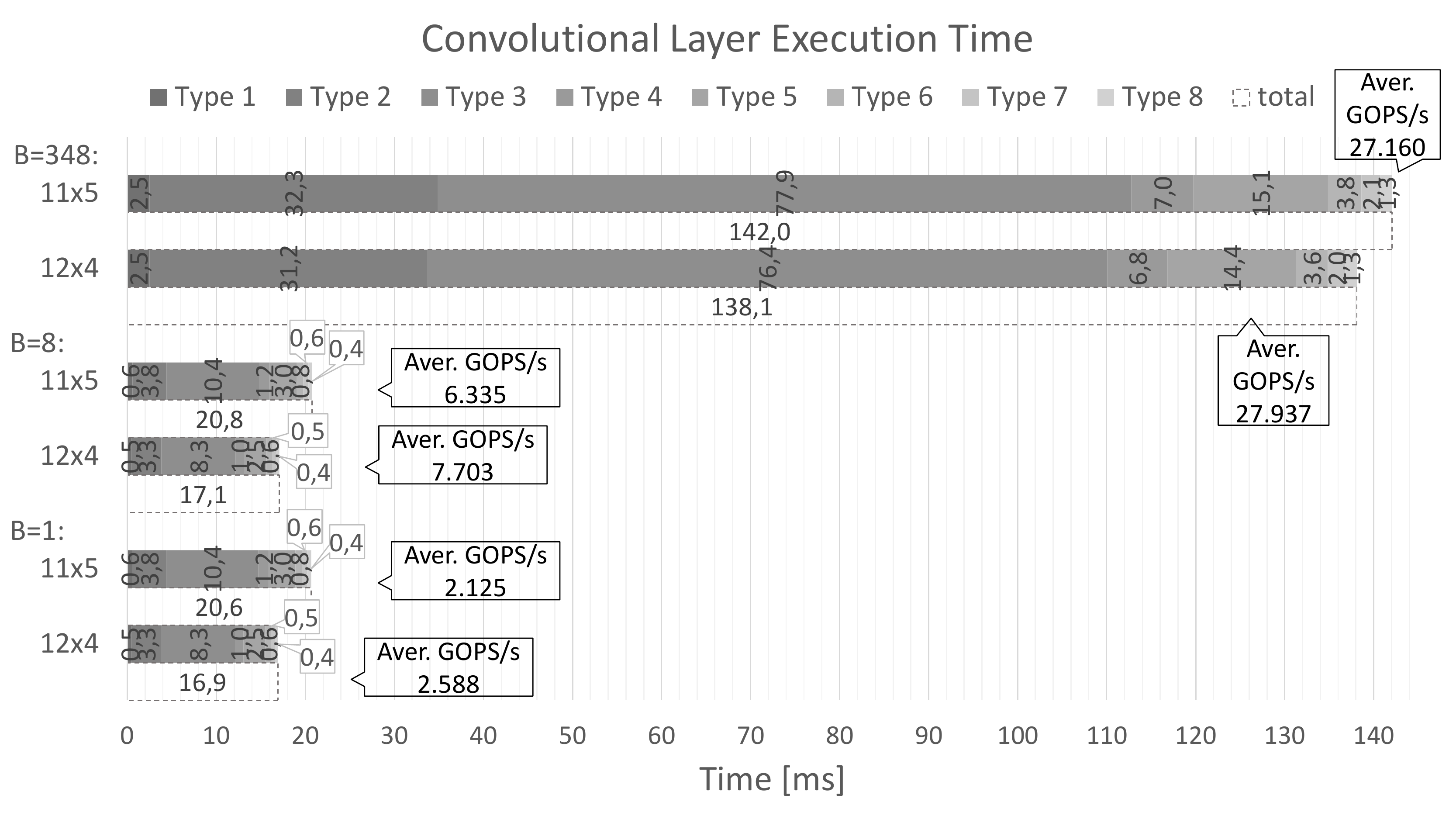}
    \caption{Execution Time comparison on ECG \cite{ecg} for NEURAghe 12x4 and 11x5 MAC matrix configurations}
    \label{fig:time_compar_zed}
\end{figure}

Figure \ref{fig:effic_compar_zed} shows the performance achieved on this benchmark by configurations implemented on the Zedboard, while Figure \ref{fig:time_compar_zed} shows execution times. 

As expected, without sample batching, performance is significantly bandwidth-limited. In minimum latency mode ($B=1$ in the Figure), efficiency is very low for every layer, and consequently on the whole network. DSPs have very long idle times and actual performance is very far from the peak. Execution time is around 17 ms with the 12$\times$4 matrix, and a bit higher with the 11$\times$5. 

To design the maximum throughput mode, we have iteratively tested different batch size values, to identify which value saturates the benefits achievable with sample batching. For this benchmark, such value is B=348, corresponding to a latency of around 1.2$s$. In this case, both configurations can operate very near to the peak performance, with an average efficiency of around 0.9. Some layers, especially Type 1, are still less efficient, however, their contribution to the overall execution time is limited. In maximum throughput mode, each network execution takes around 140 ms, producing in output classification of 348 sample consecutive sample sequences, corresponding to one sample every 0.4$ms$. 

Finally, the real-time constraint requires to process one sequence in less than 3.3 $ms$. To design the real-time operating mode, by exploring the batch size, we identified $B=8$, to be the lowest value enabling that respects such requirement. Execution time is around 17 $ms$, corresponding to around 2 $ms$ per sample.

Since 12$\times$4 is more efficient than 11$\times$5 in all modes, in this case higher frequency is more important than the number of MACs, due to better utilization with the specific layer characteristics (the matrix is partially unused in the final operations of a convolution when the number of output features is not a multiple of 5). 

%On the other hand this configuration lacks in terms of efficiency with respect to peak $GOPs/s$ reachable by NEURAghe, as already shown in Figure \ref{fig:roofline}.
%
%Low efficiency for this configuration is due to the huge weights transfer overhead with respect to the very short activation load time and execution time. So, despite the double-buffered scheduling strategy, transfer and computation phases hardly overlap.
%
%Instead, by applying the \textit{Batch processing} strategy mentioned in \ref{batch proc}, the architecture can get performances very close to the peak (rightmost column of Figure \ref{fig:effic_compar_zed}).
%
%If the target application has no latency constraint it is possible to work with much more activation samples as input for every layer. In particular when the execution time reaches and surpasses transfer times the architecture can get performances very close to the peak (lower section of Table \ref{tab:CL_perf}).
%About this, the chart at the bottom of the section shows the trend of the efficiency for every convolutional layer with respect to the \emph{batch size}. As already said, increasing \emph{batch size} leads to a better operational complexity.
%
\begin{table}[h!]
    \caption{Convolutional Layer characteristics for Network ECG \cite{ecg}}
    \centering
    \scriptsize
    
    {\def\arraystretch{1} \begin{tabular}{*{1}{c}||*{1}{c}|*{1}{c}|*{1}{c}|*{1}{c}|*{1}{c}|*{1}{c}|*{1}{c}|*{1}{c}}
 
                                                & type    & type    & type   & type  & type  & type   & type   & type  \\
                                                &   1     &  2      &   3    &   4   &   5   &   6    &    7   &  8     \\
    \hline
                             $input features$            &  1      &   320   & 256    & 256   & 128   & 128    & 128    & 64  \\
                             $output features$             &  320    &   256   & 256    & 128   & 128   & 128    & 64     & 64  \\
                             $Kernel\_size$     &  24     &   16    & 16     & 8     & 8     & 8      & 8      & 8  \\
                             $dilation\ rate$   &  1      &   2     & 4      & 4     & 6     & 8      & 8      & 8  \\
    \end{tabular}}
    \label{tab:CL_charact}
\end{table}
%
%As a third case it can be considered that for which the latency constraint is not as tightening as for the first case and it is possible to increase the efficiency without compromising performance from the point of view of the execution time.

%In particular it is possible to do interesting assumption about a real time operating mode by referring to the sample rate mentioned in the paper, that is $300\ Hz$.
%

\begin{figure}[h]
    \centering
    \includegraphics[width = \columnwidth]{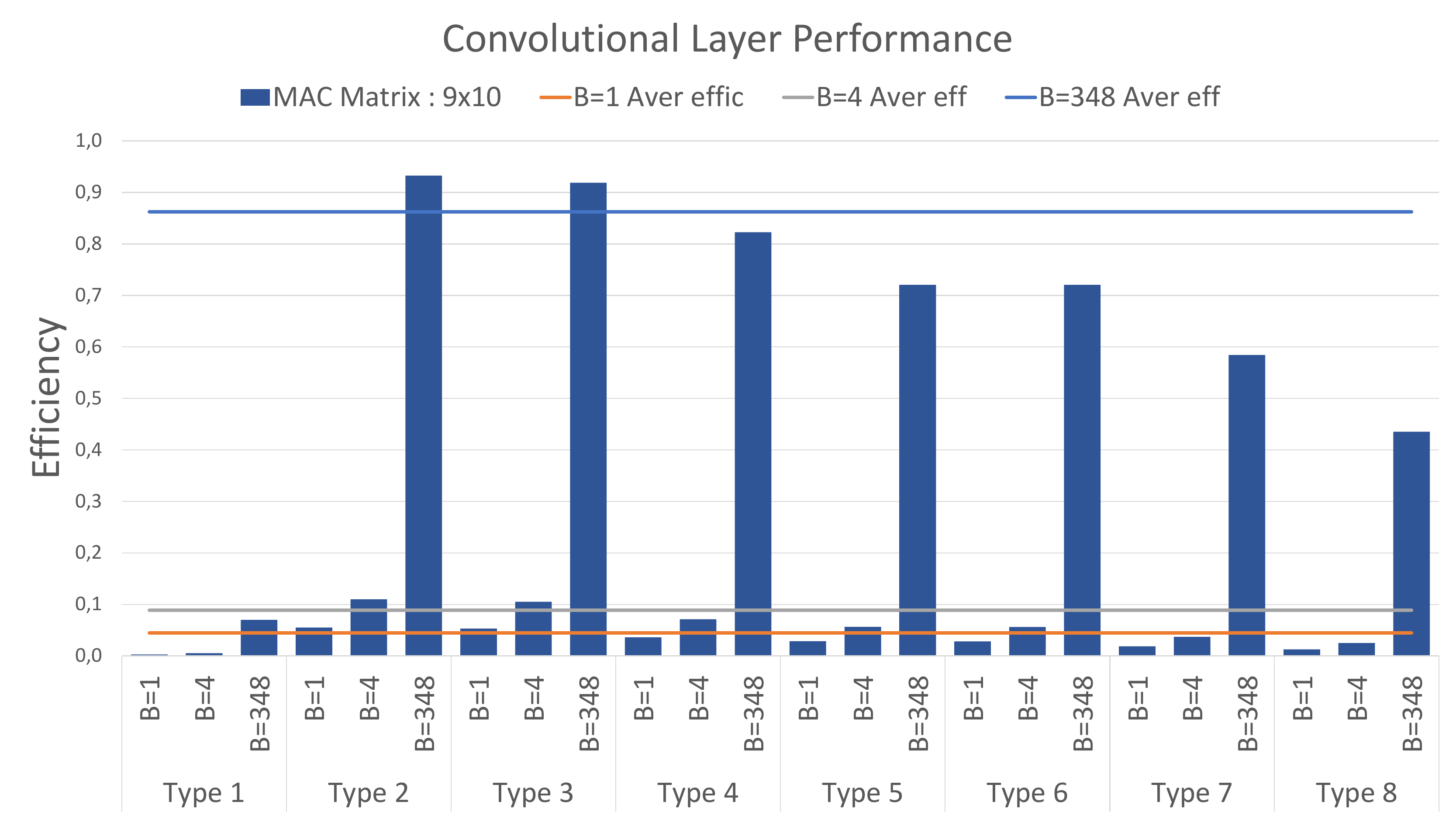}
    \caption{Efficiency on ECG \cite{ecg} for NEURAghe 9x10 MAC matrix configuration in ZU3EG}
    \label{fig:effic_compar_ultra}
\end{figure}

\begin{figure}[h]
    \centering
    \includegraphics[width = \columnwidth]{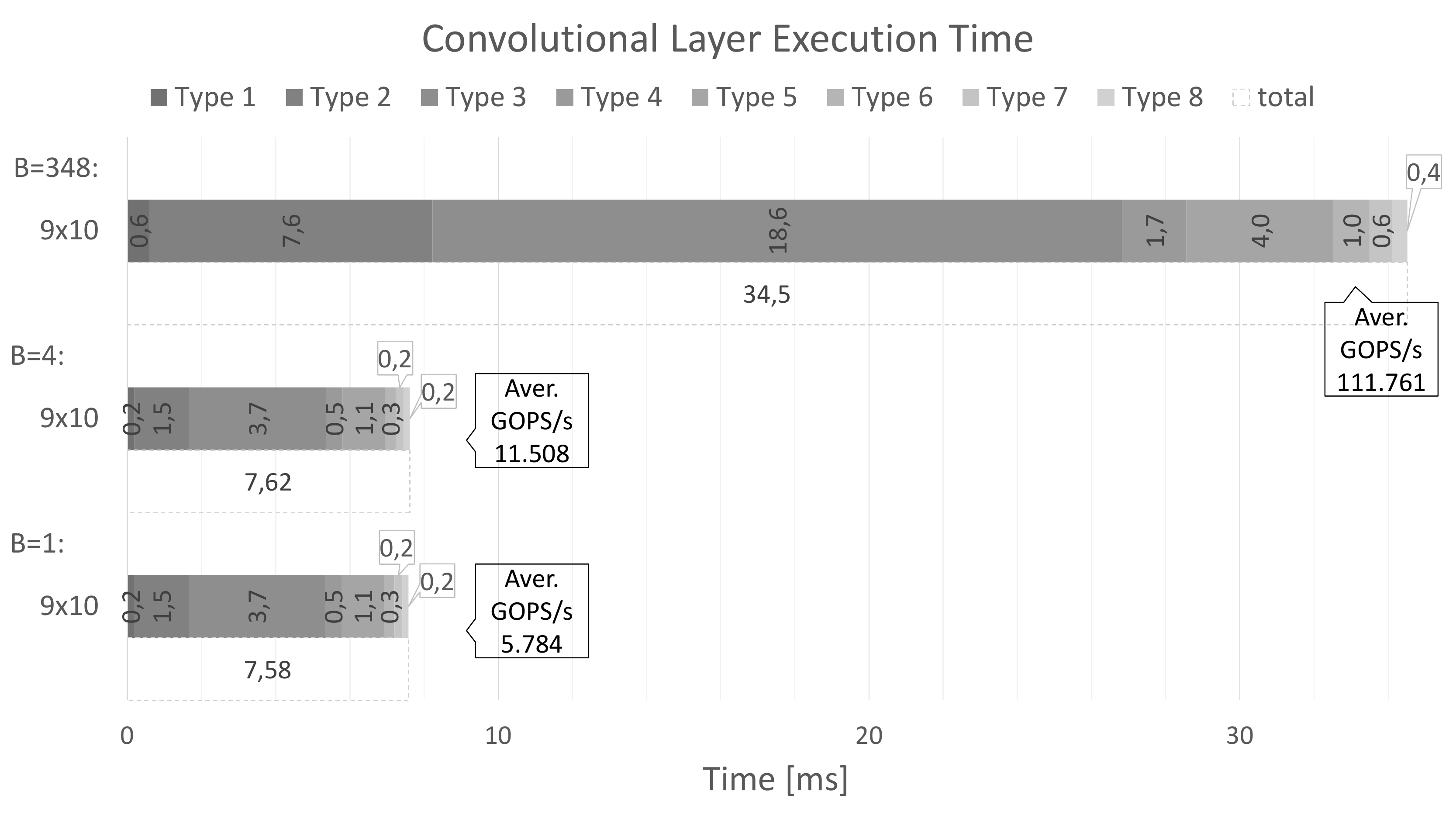}
    \caption{Execution Time on ECG \cite{ecg} for NEURAghe 9x10 MAC matrix configuration in ZU3EG}
    \label{fig:time_compar_ultra}
\end{figure}

%
%As shown in Figures \ref{fig:effic_compar_zed} and \ref{fig:effic_compar_ultra} by buffering a small amount of samples, that is increasing the $batch\ size$ parameter by $8\ samples$, before feeding the accelerator, it is possible to substantially gain in efficiency without loss in performance with respect to the \textit{latency constrained} configuration.
%Moreover, having an initial buffering of $8\ samples$ means that the accelerator has a new batch every $26.67\ ms$ which is enough to complete an end-to-end computation for this network.
The results are similar on the Ultra96 board, with increased performance, obviously, due to the higher clock frequency and the higher number of SoP modules. Efficiency and execution time are shown respectively in Figure   \ref{fig:effic_compar_ultra}  and in \ref{fig:time_compar_ultra}. Using minimum latency, efficiency is very low due to the bandwidth limits. Using $B=348$ we have around 0.9 efficiency corresponding to around 0.1 $ms$ per sample. To respect real-time constraint with minimum latency we can use $B=4$.

%\subfile{Efficiency(batch_size)}

\subsection{Res-TCN use-case}
This use case considers execution of the network presented in \cite{3D_skeleton}, made of stacked residual units, which perform 1D convolutions.
Table \ref{tab:skeleton_char_perf} shows the characteristics for the different types of Convolutional Layer in the network.
\begin{table}[]
    \caption{Convolutional Layer characteristics for Res-TCN \cite{3D_skeleton}}
    \centering
    \scriptsize
    
    {\def\arraystretch{1} \begin{tabular}{*{1}{c}||*{1}{c}|*{1}{c}|*{1}{c}|*{1}{c}|*{1}{c}|*{1}{c}}
    
                    & type    & type   & type & type  & type  & type  \\
                    &   1     &  2     &   3  &   4   &   5   &   6   \\
    \hline
                $ input features  $ & 150 & 64 & 64   & 128   & 128   & 256   \\
                $ output features $ & 64  & 64 & 128  & 128   & 256   & 256   \\
                $ Kernel\_size   $ & 8   & 8  & 8    & 8     & 8     & 8     \\
                $ stride  $ & 1   & 1  & 2    & 1     & 2     & 1     \\
    \end{tabular}}
    \label{tab:skeleton_char_perf}
\end{table}

\begin{figure}[h]
    \subfloat[\label{fig:effic_3D_hum}]{\includegraphics[width = \columnwidth, height = 0.55\columnwidth]{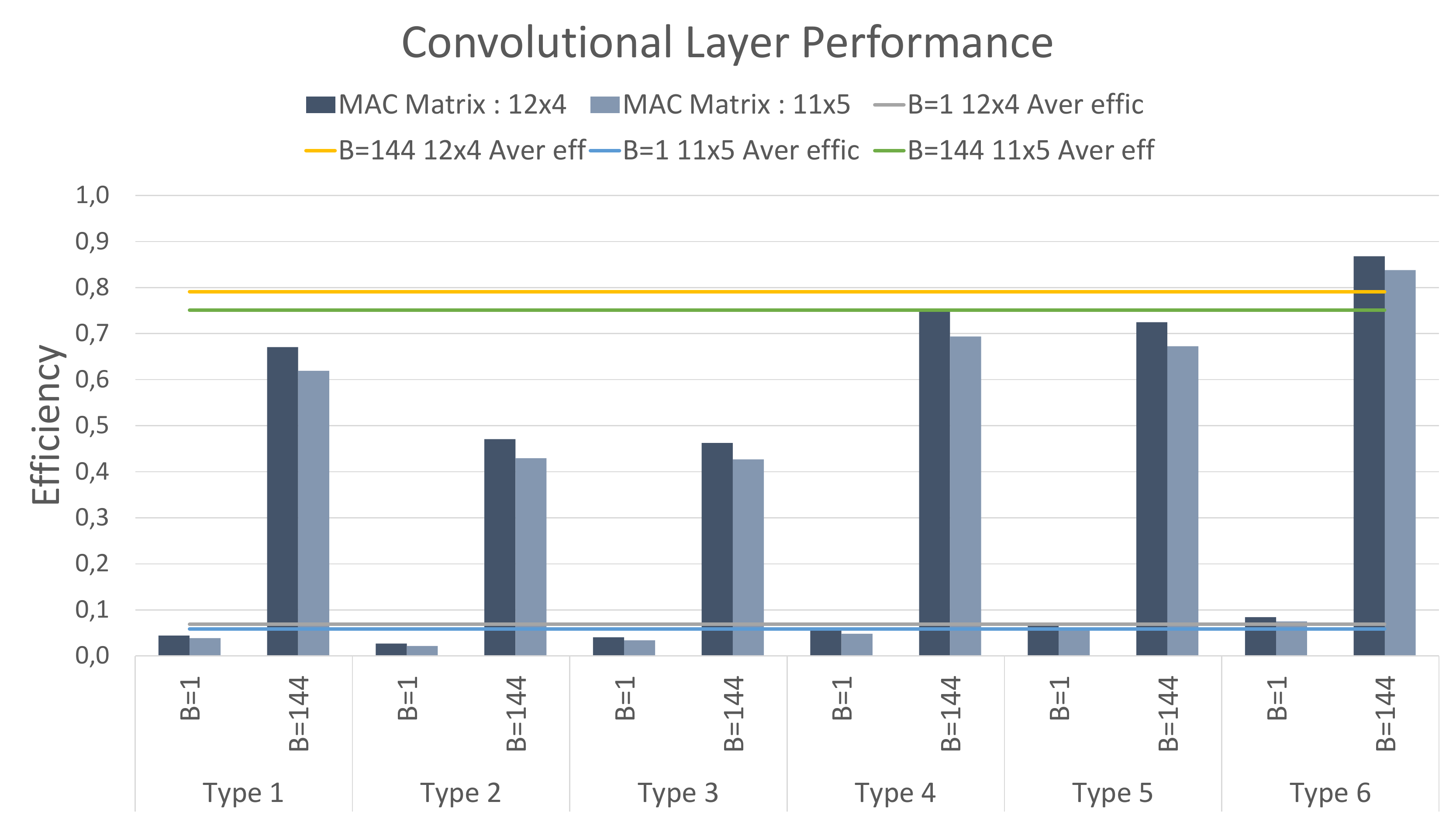}}
    \hfill
    \subfloat[\label{fig:time_3D_hum}]{\includegraphics[width = \columnwidth, height = 0.55\columnwidth]{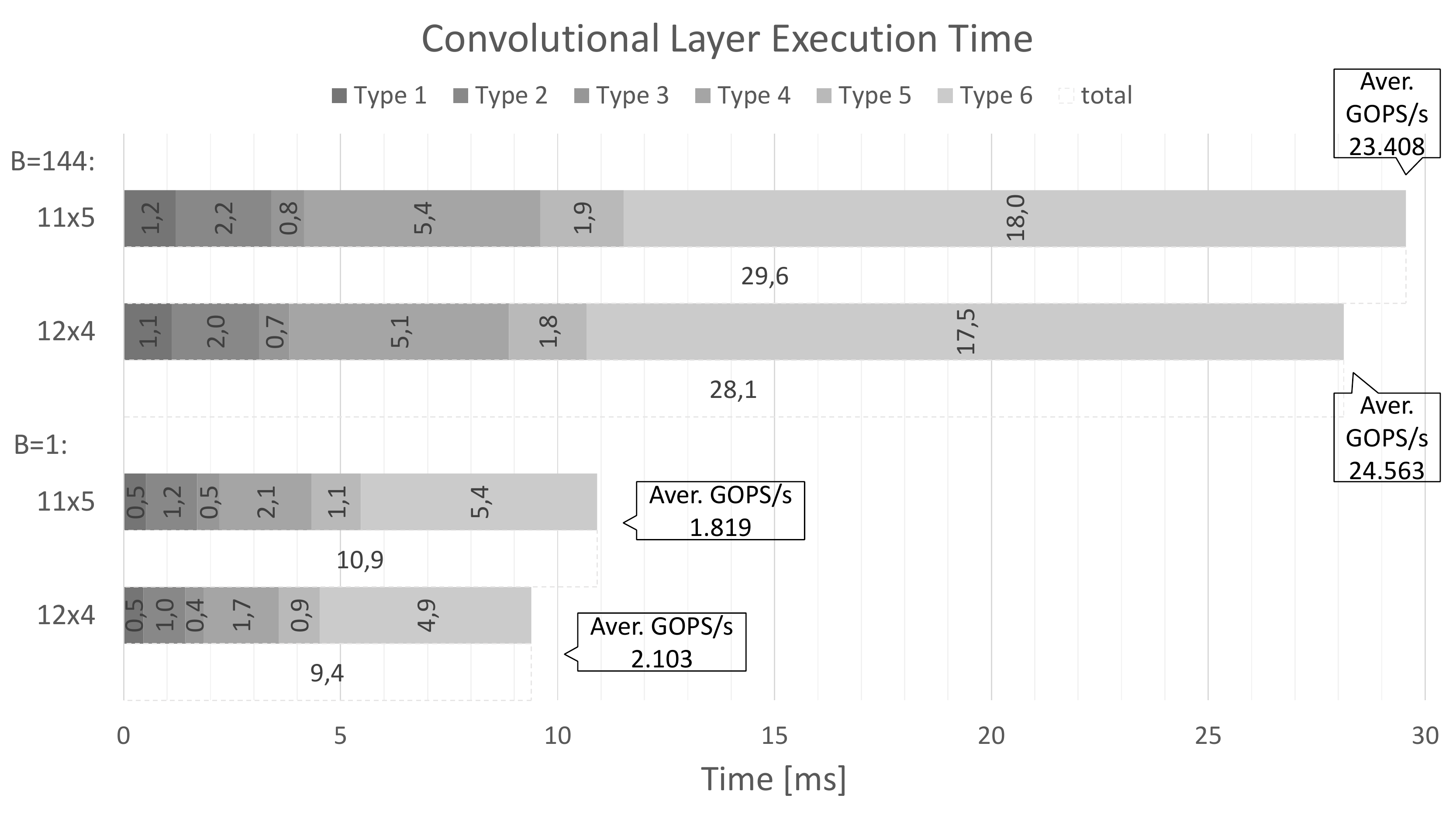}}
    \caption{Efficiency trend and Execution Time on Res-TCN \cite{3D_skeleton} for different batch sizes for NEURAghe 12x4 and 11x5 matrix configuration in Z-7020 SoC}
    \label{fig:effic_time_3D_hum}
\end{figure}

\begin{figure}[h]
    \subfloat[\label{fig:effic_3D_hum_ultra}]{\includegraphics[width = \columnwidth, height = 0.5\columnwidth]{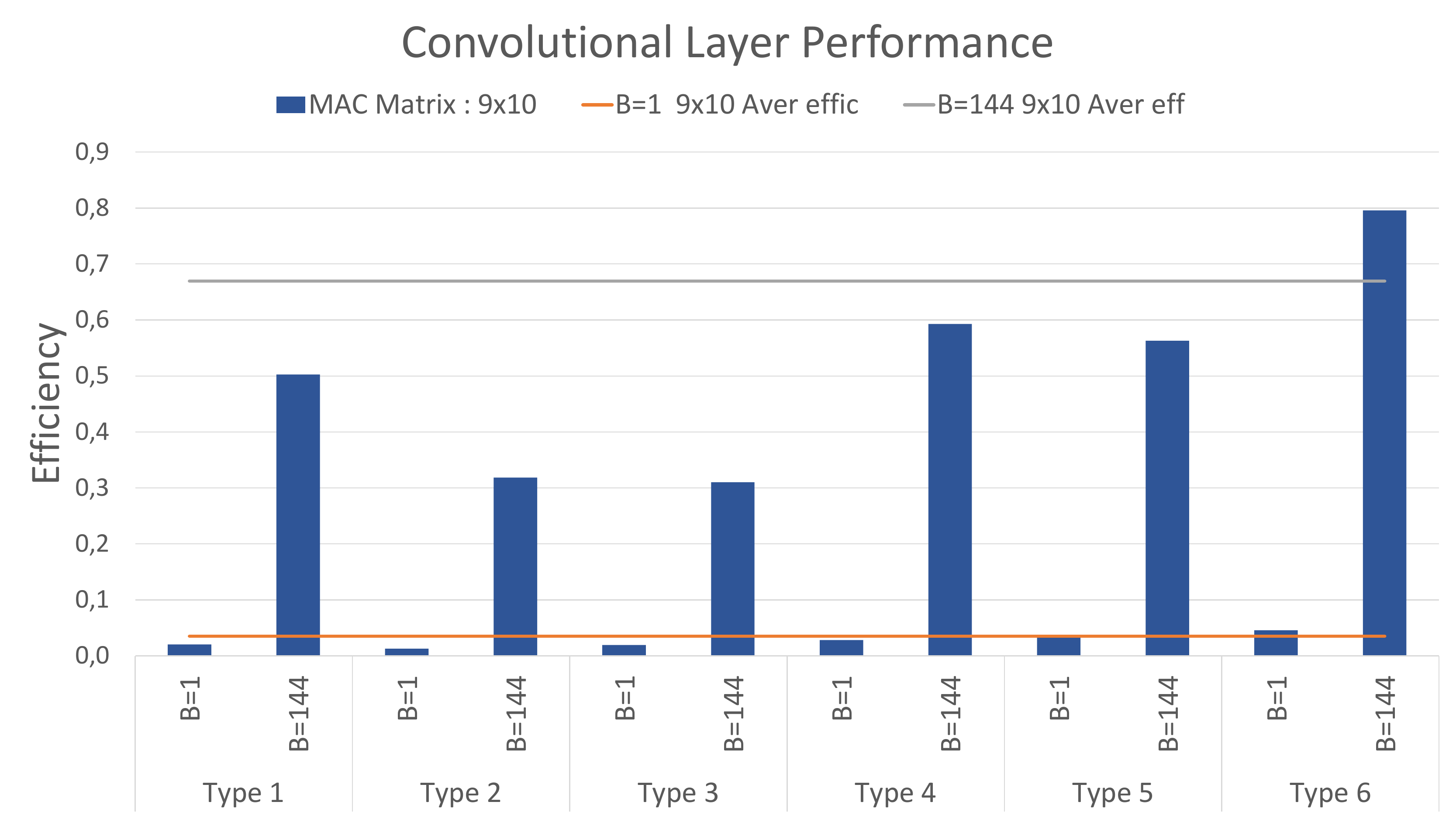}}
    \hfill
    \subfloat[\label{fig:time_3D_hum_ultra}]{\includegraphics[width = \columnwidth, height = 0.5\columnwidth]{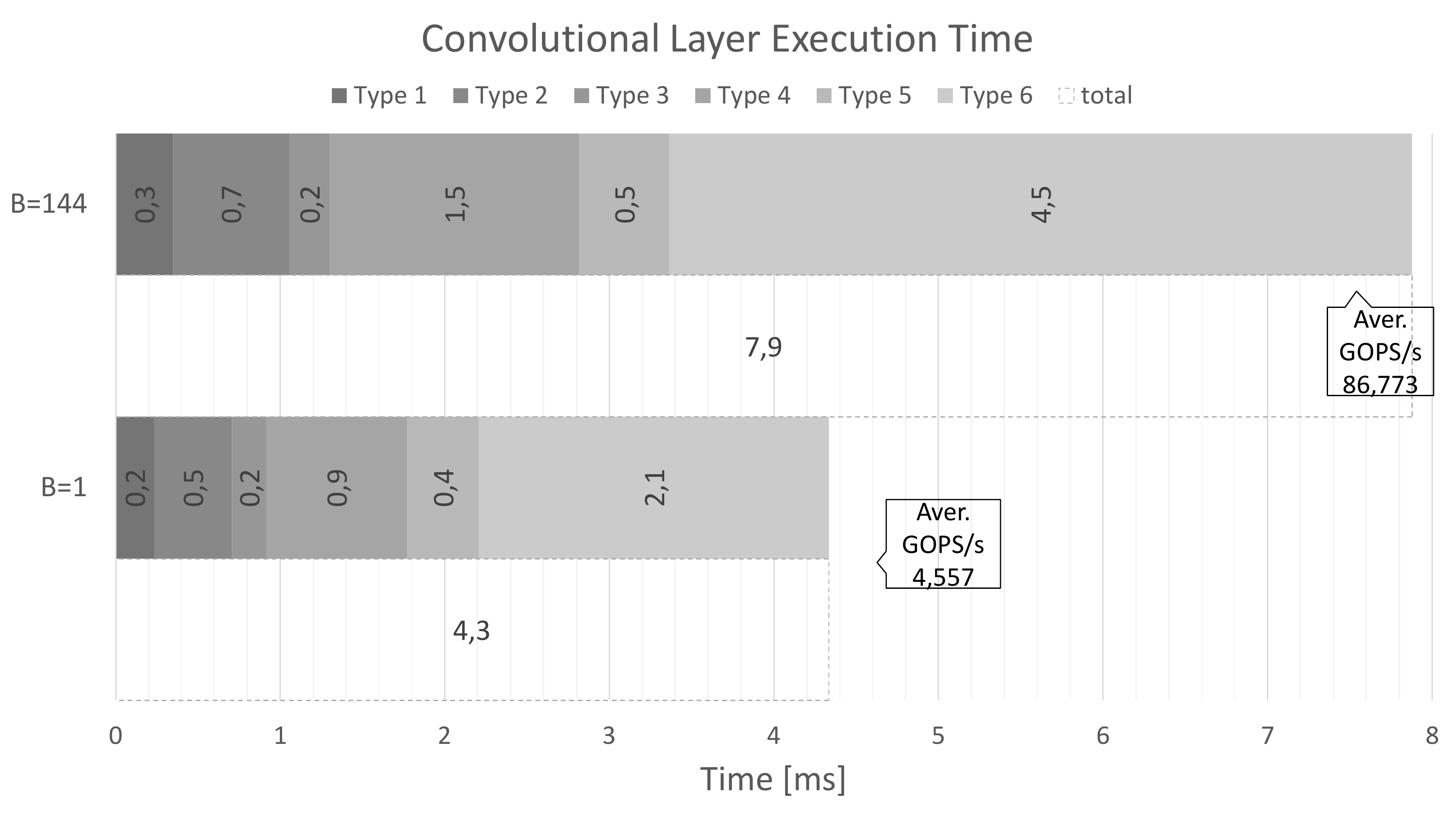}}
    \caption{Efficiency trend and Execution Time on Res-TCN \cite{3D_skeleton} for different batch sizes for NEURAghe 9x10 matrix configuration in ZU3EG SoC}
    \label{fig:effic_time_3D_hum_ultra}
\end{figure}

Figure \ref{fig:effic_time_3D_hum} shows efficiency and execution time on this benchmark, highlighting contributions of the single layers. 
Behaviour is similar to the ECG use case on both boards. Executing the processing after each sample ($B=1$), the efficiency is very limited for all the layers. Increasing $B$, the accelerator provides much better throughput. The layers executed less efficiently are the early layers, especially the first type, that pays for a significant underutilization of the MAC Matrix, due to the low number of input features. However, its contribution to the overall efficiency is marginal. Layers that show longer execution time are placed at the end of the TCN graph. These layers account for the highest contribution to the overall workload and, as shown in Figure \ref{fig:time_3D_hum} and \ref{fig:time_3D_hum_ultra}, are executed quite efficiently when sample batching is used ($B=144$). Sample batching is more effective on the Zedboard, that integrates a MAC Matrix with less input and output ports, thus requires longer runs to complete convolution, which reduces the impact of data-transfer and accelerator warm-up overheads. 

%
%As can be seen it is possible to gain efficiency by increasing this parameter and by accepting a bit of latency.

%\subsection{Polyphonic Note Transcription with WaveNet}

%Another whole category of Temporal Convolutional Networks are those inspired by $WaveNet$ (Van Den Oord et al. \cite{wavenet}) a probabilistic and autoregressive model for raw audio generation, in which every new predicted sample is conditioned on all previous ones.
%The model is composed by different stacked $residual\ blocks$ connected together by means of residual connections and producing skip connections used to speed up convergence.
%Every residual block contains a gated dilated convolutional block followed by a $1\times 1$ convolution useful for channel normalization before summing together with the input to produce the output for the residual connection.
%The gated activation units applied to the dilated convolution result are usually a $tanh(\cdot)$ and a \emph{sigmoid function} $\sigma(\cdot)$. After these, an element-wise multiplication feeds the input to the $1\times 1$ convolution.
%The $1\times 1$ output feeds also skip connections.

\subsection{WN-PNT use-case}
This case considers a network derived from WaveNet (Van Den Oord et al. \cite{wavenet}) for Polyphonic Note Transcription of Time-Domain Audio Signal (\cite{polyWaveNet}). It is made of $20$ stacked residual blocks composed by a $1\times 1$ skip connection and a Dilated Convolutional block of $128$ channels each. \emph{Filter Size} is $2$ and the \emph{dilation factor} grows like $2^{k}$ where the residual block index $k \in \{0,1,2 ... 9,0,1,2 ... 9\}$.

%\begin{figure}[h]
%    \subfloat[\label{fig:effic_poliph}]{\includegraphics[width = \columnwidth, height = 0.55\columnwidth]{img/effic_poliph.pdf}}
%    \hfill
%    \subfloat[\label{fig:time_poliph}]{\includegraphics[width = \columnwidth, height = 0.55\columnwidth]{img/time_poliph.pdf}}
%    \caption{Efficiency trend and Execution Time on Polyph on WaveNet \cite{polyWaveNet} for different batch sizes for NEURAghe 12x4 and 11x5 matrix configuration in Z-7020}
%    \label{fig:effic_time_poliph}
%\end{figure}

\begin{figure}[h]
    \subfloat[\label{fig:effic_poliph_ultra}]{\includegraphics[width = \columnwidth, height = 0.5\columnwidth]{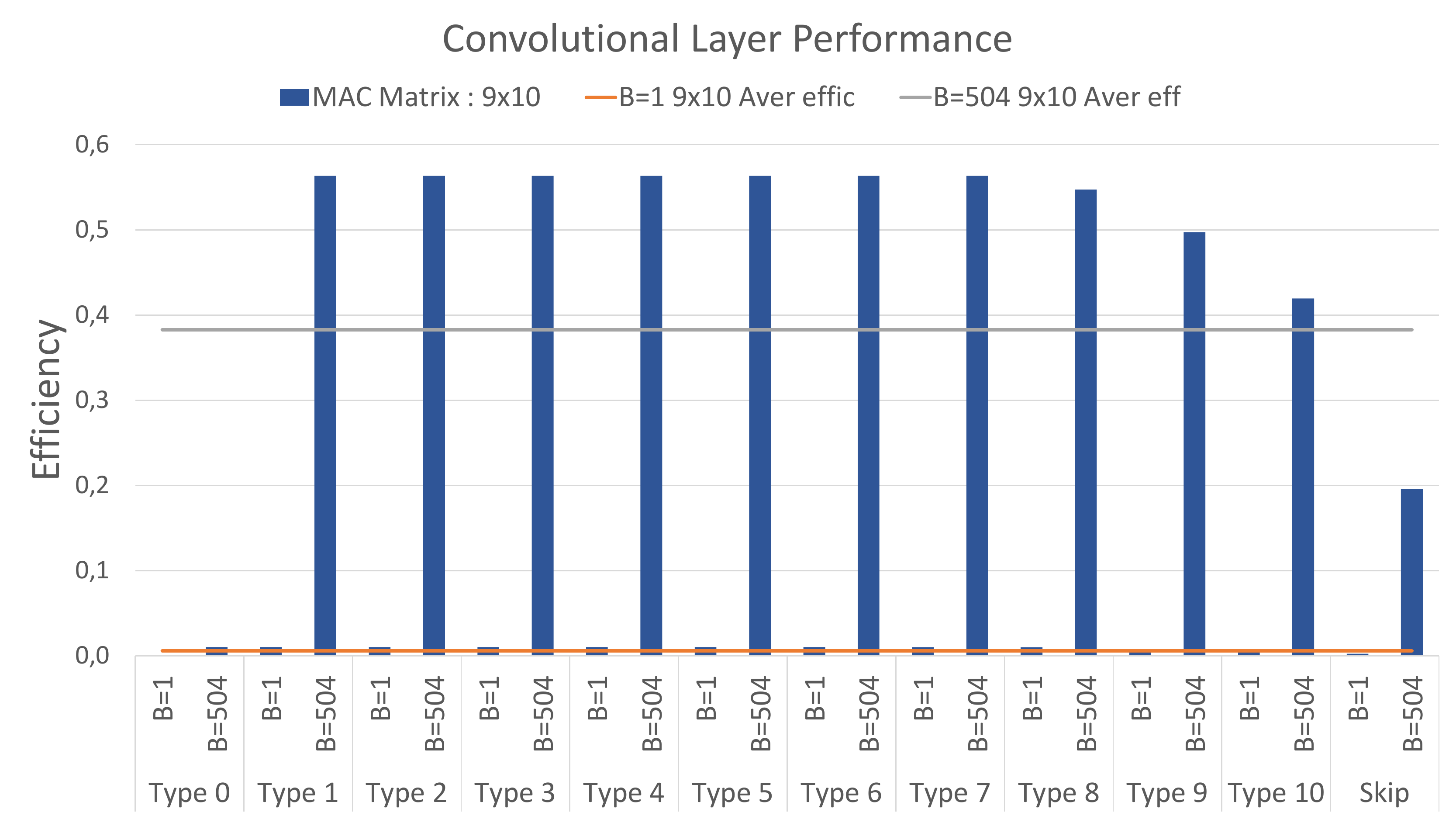}}
    \hfill
    \subfloat[\label{fig:time_poliph_ultra}]{\includegraphics[width = \columnwidth, height = 0.5\columnwidth]{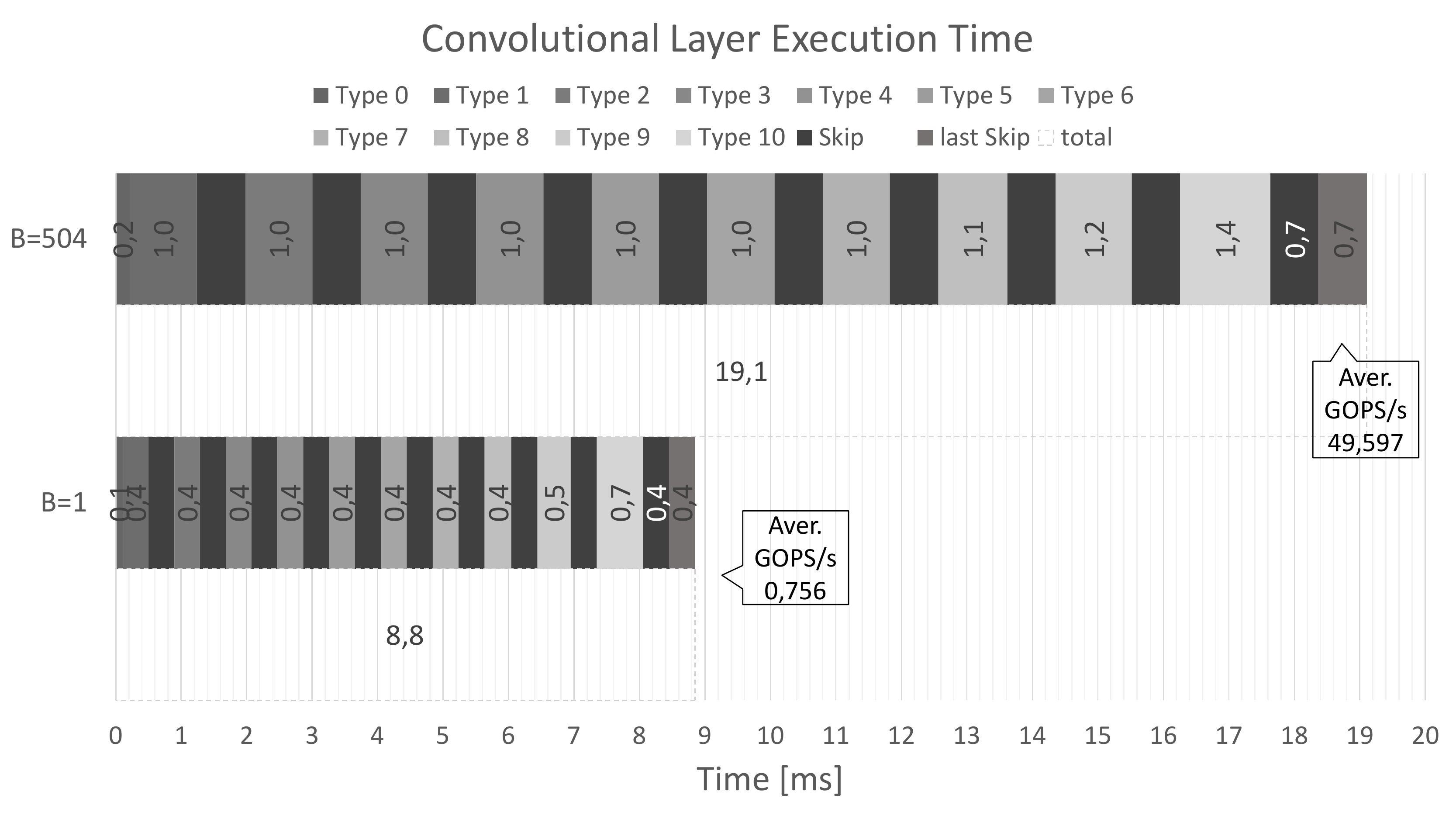}}
    \caption{Efficiency trend and Execution Time on WN-TCN \cite{polyWaveNet} for different batch sizes for NEURAghe 9x10 matrix configuration in ZU3EG}
    \label{fig:effic_time_poliph}
\end{figure}

This case is especially challenging, due to the small size of kernels ($kernel\_size=2$ in most layers) and because the sample acquisition frequency is $16KHz$, posing hard constraints about real-time execution. Thus we focus on the Ultra-96 board, relying on its higher clock frequency to achieve the required throughput. 
As may be noticed in Figure \ref{fig:effic_poliph_ultra}, in general, this use case is executed less efficiently on the platform, none of the layers reaches more than 0.56 efficiency. This is because the execution of the actual convolution kernels takes only two cycles and thus hardly overlaps with input data transfers. Batch size can be used to increase the reuse of weights, once they are loaded to the weight memory region, but, at the same time, has an impact on the duration of input and output transfers. To compensate this issue, considering that the total size of the local receptive fields in the network allows for the continuous storage on the on-chip memory for almost all layers, (unless for types 8, 9 and 10), we have slightly modified the scheduling described in Figure \ref{fig:scheduling}. In this use-case, we transfer to/from DDR only new output/input samples, while keeping the rest of the local receptive fields in the on-chip memory. As shown in Figure \ref{fig:time_poliph_ultra}, using $B=504$ we can process 504 samples in around $19 ms$, thus respecting the real-time constraint posed by the use-case.

\subsection{Assessment of hardware vs. software speed-up}
\begin{figure}[h]
    \subfloat[\label{fig:neu_a53_time}]{\includegraphics[width = \columnwidth, height = 0.5\columnwidth]{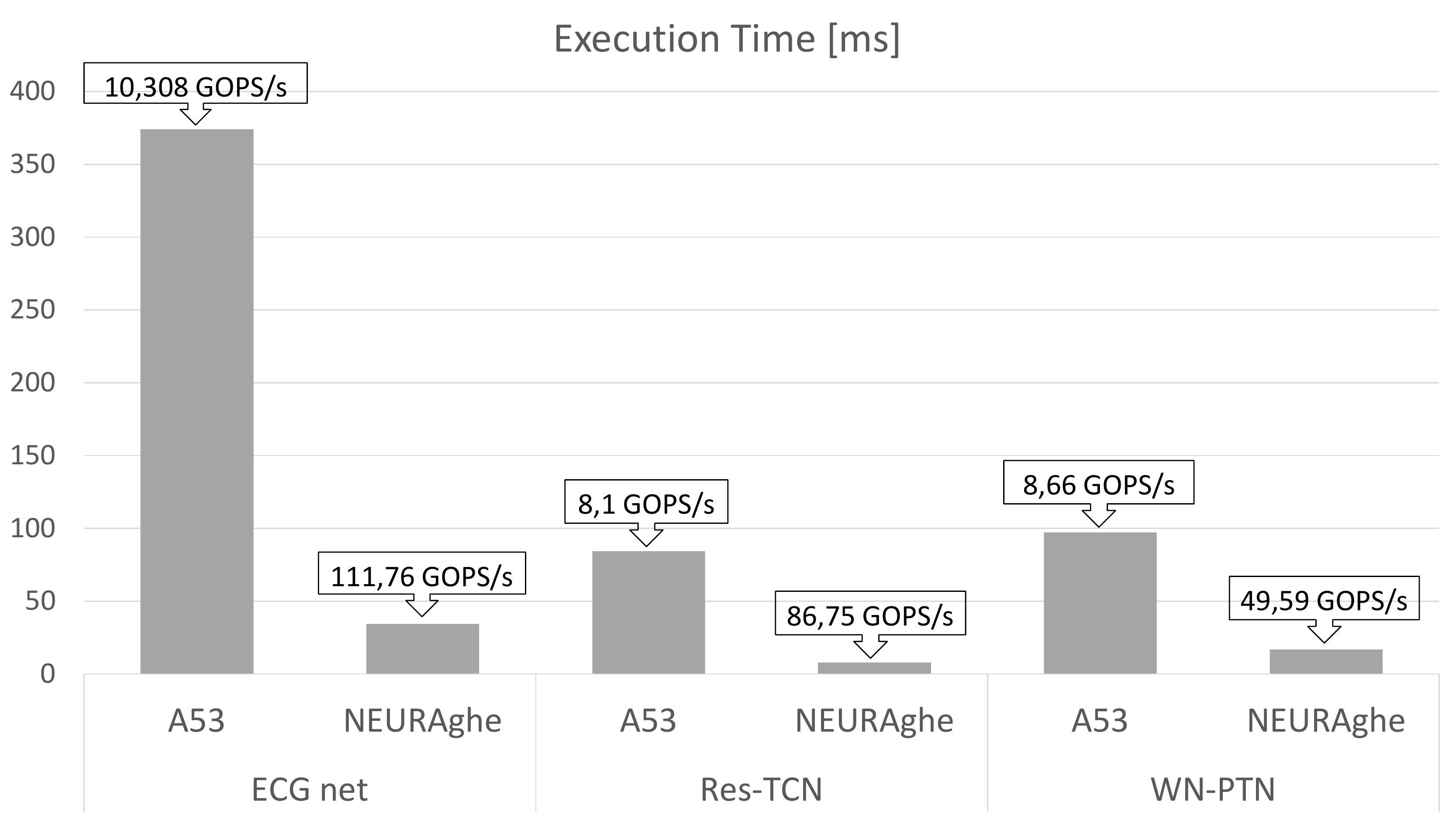}}
    \hfill
    \subfloat[\label{fig:neu_a53_watt}]{\includegraphics[width = \columnwidth, height = 0.5\columnwidth]{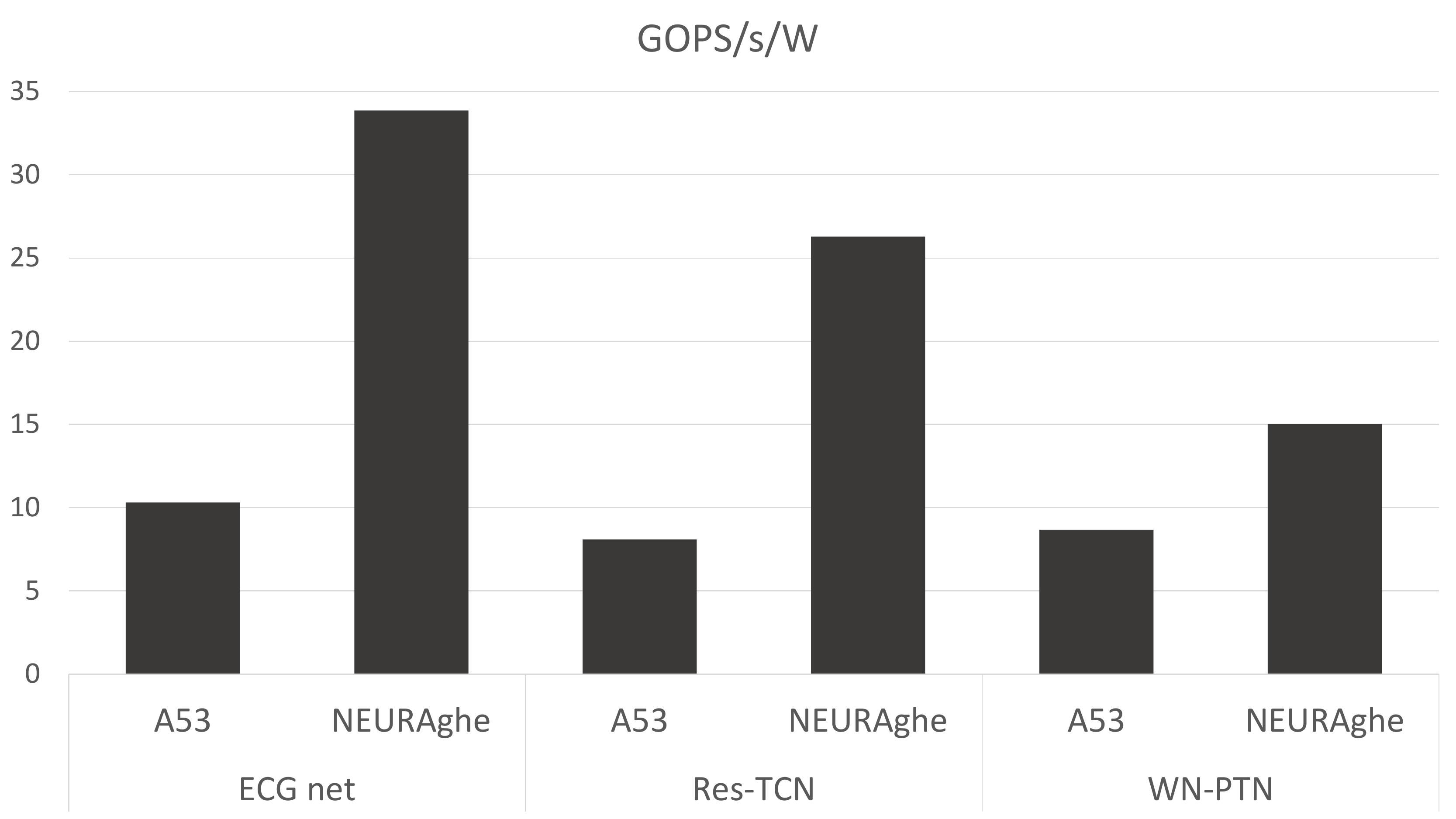}}
    \caption{Execution time and power efficiency comparison  between software execution on a Cortex-A53 quad core and NEURAghe (ultra-96).}
    \label{fig:neu_a53}
\end{figure}

In order to evaluate the benefits obtained by on-FPGA acceleration, we have compared the execution of the convolution-related load of the three benchmarks on NEURAghe, with the execution on an A53 quad-core ARM processor, using the ARM Compute Library optimized functions \footnote{Since ARM-CL does not support dilated 1D convolution, for the sake of comparison, we have evaluated execution time on the A53 on analogous convolution layers, without dilation. Considering that in dilated convolutions input samples are not adjacent in memory, the exploitation of the vector processing in the ARM-CL could be suboptimal, thus the execution time reported in Figure \ref{fig:neu_a53_time} for the A53 could be underestimated. Our own software implementation (unoptimized) of a dilated convolution performs one order of magnitude slower than what reported for ARM-CL.}. As may be noticed, NEURAghe provides up to 10.8x, 10.7x and 5.7x speed-up on the three reference benchmarks, and considering the power consumption of the two platforms (around $0.9\ W$ for the A53 cores and around $3.3\ W$ for NEURAghe), PL-based acceleration provides improvement, in terms of power efficiency, corresponding to 33,8 GOPS/s/W , 26,3 GOPS/s/W and 15 GOPS/s/W respectively.

\subsection{Comparison to other APSoC based accelerators}
As mentioned, literature is lacking TCN evaluations on FPGA-based accelerators. The only attempt available is \cite{waveFPGA}, which is strictly customized for a autoregressive TCN and is implemented on a high-end board non usable in the embedded domain. However, when designing the TCN-supporting version of NEURAghe, we have considered compatibility with 2D convolutions for classic CNN acceleration. Moreover, the improved flexibity of the architecture, in terms of kernel size and stride, allows for some improvements with respect to other approaches in the literature targeting low-to-mid All-programmable SoCs. 
In Table \ref{tab:comp_previous}
we report comparative results with state-of-the-art on two well known networks for image classification, ResNet-18 and VGG-16. We compare to other accelerator architectures, \cite{neuESL}, \cite{sharma2016high} and \cite{venieris2017latency}, that are implemented on the same kind of hardware and use the same 16-bit data precision. 
On VGG-16, that exposes quite regular kernel sizes and stride values, our work shows comparable performance with respect to the alternatives. It executes convolutions slightly faster than \cite{neuESL} and \cite{sharma2016high} and around 13\% slower than \cite{venieris2017latency}. On ResNet-18, which exposes more  variable $kernel\ sizes$ and $strides$, we can reduce a lot of overheads that must be payed by more \emph{static} architectures, executing the whole convolution workload 40\% faster than \cite{neuESL}.
\begin{table}[h]
    \caption{Comparison between this work, previous version (\cite{neuESL}) and other works on ResNet-18 and VGG-16. Xilinx Zynq Z-7020}
    \centering
    \scriptsize
     {\def\arraystretch{1} \begin{tabular}{*{1}{c}|*{1}{c}*{1}{c|}*{1}{c}*{1}{c}*{1}{c}*{1}{c}}
                & \textbf{This}   &     & \textbf{This}  & &   &    \\
                & \textbf{Work}   & \cite{neuESL}  & \textbf{Work} & \cite{neuESL} &  \cite{venieris2017latency} &  \cite{sharma2016high} \\
    \hline
                & \multicolumn{2}{c|}{ResNet-18} & \multicolumn{4}{c}{VGG-16}   \\
    \hline
            Xilinx Zynq SoC & \textbf{Z-7020} & Z-7020   & \textbf{Z-7020} & Z-7020 & Z-7020 & Z-7020 \\
            Freq. [MHz]     & \textbf{120}    & 120      & \textbf{120}    & 120    & 125    & 150 \\
            GOPS/s          & \textbf{26,5}  & 16,1    & \textbf{42,62}  & 42.48  & 48.53  & 31.38 \\

    \end{tabular}}
    \label{tab:comp_previous}
\end{table}

\section{Conclusions and future work}\label{sec:concl}

In this work, we have presented an accelerator architecture supporting Temporal Convolutional Networks, implemented on FPGA-based SoCs. We have presented the accelerator features serving this specific computing pattern, such as the capability of supporting arbitrary \emph{kernel\_size}, \emph{dilation rate} and \emph{stride values} without overhead.
We have also shown how data transfers from-to off-chip memory can be managed in TCNs and how performances improve by changing the computational paradigm, going from a \emph{latency constrained} approach to a \emph{batched} approach, that trades latency for throughput.
We applied this method to three notable TCNs and using two different SoCs as a target, analyzing throughput, latency and efficiency figures, and the capabilities of the system to respect real-time constraints.
Results show that using sample batching, we can achieve efficiency up to 0.96, 0.86 and 0.57 on the three use-cases. In the two use-cases with real-time requirements, an adequate sample batching can be used to achieve sufficient throughput to timely process all input samples. 
We also compared the execution of the use-cases on a Cortex A53 quad-core, to evaluate the achievable speedup, noticing up to 10x execution time reduction, with an improvement in power efficiency that ranges from 2x to 3.5x depending on the benchmark. 
Finally, we validated the compatibility of proposed architectural solutions with state-of-the-art CNNs used in the image processing domain, by direct comparison with previous work on APSoC-based accelerator, showing similar performance with respect to CNN-targeting alternatives when regular network topologies are targeted and 40\% improvement when targeting more irregular patterns. 
%The next step can be exploring different architectural configurations suitable both for different target devices and different TCNs in order to find the solution that adapts best.
\balance
\bibliographystyle{IEEEtran}
\bibliography{IEEEabrv,neuraghe}

\end{document}